\documentclass[
prx,
superscriptaddress
]{revtex4-1}

\usepackage{float}
\usepackage{bbm}
\usepackage{epsfig}
\usepackage{epstopdf}
\usepackage{graphicx, subfigure}
\usepackage{pgfplots}
\usetikzlibrary{patterns}
\usepgfplotslibrary{groupplots}
\usepackage{amssymb}
\usepackage{amsmath}
\usepackage{scalefnt}
\usepackage{color}
\usepackage[title]{appendix}
\usepackage{hyperref}
\usepackage{qcircuit}
\usepackage[capitalize]{cleveref}
\usepackage{braket,mleftright}
\usepackage{physics}
\usepackage{import}
\usepackage{dcolumn}
\usepackage{csquotes}
\usepackage{tikz}
\usepackage{comment}

\floatstyle{plaintop}
\restylefloat{table}
\begin{document}

\title{Teaching quantum information science to high-school and early undergraduate students}

\author{Sophia E. Economou}
\email{economou@vt.edu}
\affiliation{Department of Physics, Virginia Tech, Blacksburg, VA 24061, U.S.A}
\author{Terry Rudolph}
\affiliation{Department of Physics, Imperial College London, London SW7 2AZ, UK}
\author{Edwin Barnes}
\affiliation{Department of Physics, Virginia Tech, Blacksburg, VA 24061, U.S.A}
    
\date{\today}

\begin{abstract}
We present a simple, accessible, yet rigorous outreach/educational program focused on quantum information science and technology for high-school and early undergraduate students. This program allows students to perform meaningful hands-on calculations with quantum circuits and algorithms, without requiring knowledge of advanced mathematics. A combination of pen-and-paper exercises and IBM Q simulations helps students understand the structure of quantum gates and circuits, as well as the principles of superposition, entanglement, and measurement in QM. 
\end{abstract}

\maketitle

\section{Introduction}

There has been a huge expansion in the field of Quantum Information Science and Technology (QIST) over the last few years, with large investments by industry and governments worldwide. As the field expands, so do workforce needs and the public's exposure to QIST, at least at a superficial level. Students read about quantum computing and related technologies in popular science articles and become curious and eager to learn more. Yet, there is a barrier for them to enter these fields, as typically they have to go through a Physics (or related field) curriculum, and even then they only get to learn and use the full mathematical machinery of quantum mechanics (QM) in their senior, or at best junior, year. This is because, traditionally, students first spend substantial time learning to solve the Schr\"odinger equation in position space before they see finite Hilbert space problems such as a spin in a magnetic field. Some books \cite{Feynman-lectures, Townsend-book,McIntyre-book,Wilcox-book} start instead from finite Hilbert spaces, which makes the subject more accessible, as in this case the main prerequisite is linear algebra. In fact, one can learn quantum information without having taken a QM course, and there exist textbooks which take this approach, such as, e.g., the excellent book by Mermin on quantum computing \cite{Mermin-book}. Refs.~\cite{Dur2014,Dur2016,Fermilab-paper,Shoshany2018} present quantum computing high-school modules that also start from finite Hilbert spaces and also assume knowledge of linear algebra or introduce it quickly in the beginning of the module. This can be a barrier, though, since linear algebra is not typically covered in a standard high-school curriculum (at least not in the United States). An ambitious multimedia-based MOOC has been developed to teach QM to non-scientists~\cite{Freericks2019}. However, this still requires students to commit about a month to complete the course. In general, existing resources require either some knowledge of advanced math beyond what is covered in high school or a considerable time investment (several weeks) before they can meaningfully tackle problems and gain an actual understanding of QIST. This can limit the scope and audience of QIST outreach activities aimed at attracting younger students to STEM fields and at enhancing scientific literacy among the general public.

Here we describe an outreach program that two of us (EB, SEE) have developed under an NSF sponsored EFRI project. Our approach is in part based on a simple machinery devised by one of us (TR) in 2015 when asked to teach some classes on quantum computing at a math camp for 12-14 year-olds in the UK and later refined for a week long series of lectures at the African Institute for Mathematical Sciences in Rwanda in early 2017. These lectures were to Masters-level students whose background was in statistics and data analysis. The lecture notes from that course were turned into the book `Q is for Quantum' \cite{Rudolph-book}, which allows students who do not have any linear algebra (or other sophisticated math) background to get a substantial understanding of the basics of quantum information and perform simple calculations. A pdf copy of Part I of the book is available for free at \href{http://qisforquantum.org/}{qisforquantum.org}. We henceforth refer to the book and the formalism it introduces as QI4Q. The rest of the outreach program developed by EB and SEE uses IBM Quantum (IBM Q) Experience simulators and devices, where the students run circuits and compare the results to the pen-and-paper work they do using the QI4Q formalism. The final stage involves a quantum game developed by one of us (EB) called ``Money or Tiger". 
To summarize, the outreach program has four elements:
\begin{itemize}
\item Overview of QM and QIST (lecture)
\item Practice with formalism from `Q is for Quantum' (hands-on activity)
\item Practice with IBM Q experience (hands-on activity)
\item Solution of a game based on a quantum algorithm (hands-on activity)
\end{itemize}

We have used this approach in a few different settings. The lecture part of this program along with a few elements of the QI4Q formalism and the Money or Tiger game were presented at a few Virginia high schools by one of us (EB). These elements were then combined with IBM Q experience hands-on activities in a two-day program two of us (SEE and EB) carried out as part of one of the programs organized by Virginia Tech's Center for the Enhancement of Engineering Diversity \cite{ceed}. This particular program, called CTech$^2$ \cite{ctech2}, is a two-week camp for approximately 60 rising junior and senior female high-school students. The students stay on campus in the dorms and spend their days learning various engineering and science topics, with an emphasis on hands-on activities. The most comprehensive version of our outreach program was employed for CTech$^2$, where we held a two-day event, so this is what we will focus on below. One of us (EB) has also used some of this material to teach a quantum information module of a freshman physics course at Virginia Tech called `Highlights of Contemporary Physics', which is taken by both physics majors and non-physics majors. 

The purpose of this paper is three-fold: (a) to share with others our approach, which we have found to be successful; this material could be used by other researchers who want to develop QIST outreach activities or by high-school teachers to complement their science classes; (b) to provide a simple step-by-step guide to high-school or early college students (or even laypersons) who are interested in getting hands-on experience and some familiarity with QIST; (c) to inspire others in the QIST community to contribute by coming up with new games based on our approach. We emphasize that our approach is not intended to replace proper, linear-algebra-based treatments of QIST. Instead it is meant to offer a brief yet accessible and hands-on introduction to this field that will hopefully motivate students to pursue more in-depth university courses in the future and instill a general appreciation for QIST.

This paper is organized as follows. In Sec.~\ref{qmoverview} we describe the lecture we give on QM, where we take a more historical and `traditional' perspective. In Sec.~\ref{QI4Qformalism} we describe the lecture on QIST and the QI4Q formalism. In Sec.~\ref{ibmq}, we present the hands-on exercises the students do on IBM Q simulators and processors. In Sec.~\ref{moneyortiger} we present the ``Money or Tiger" game along with its solution and implementation on IBM Q simulators and devices. Finally, Sec.~\ref{limitations} provides a discussion on the limitations of this formalism. 

\section{Overview of quantum mechanics lecture}\label{qmoverview}

The lecture on QM mostly follows a historical perspective and is essentially a condensed version of treatments that can be found in many introductory textbooks. Although this approach may not seem especially novel, in practice we find that it has merits. Our experience is that students are able to absorb the basic concepts of QM more easily if they understand what motivated scientists to introduce these concepts in the first place. In addition, this approach provides a natural way to highlight the gaps in human knowledge that persist to this day and to show that many of the confusions students may experience in the course of the lecture are often shared by quantum researchers. This treatment also emphasizes the fact that QM is a fundamental building block of the universe and not just a special feature that enables QIST applications. The main objective of this lecture is to give students a basic understanding of superposition and probability in QM.

The lecture begins with some general statements summarizing the classical physics encountered by the students in their high school physics classes. Particular emphasis is given to the concept that everything is made of particles, and that in classical physics, everything is deterministic in the sense that if we know the initial positions and velocities of particles, we can predict their future positions and velocities. The point, of course, is to later contrast this with what happens in QM. We point out that this classical understanding of the universe persisted for 200 years, up until the first decade of the 20th century. We then stress that a short period of radical change ensued, such that by 1930 scientists no longer believed that particles even have well defined positions or velocities, and that Nature is fundamentally probabilistic. Presenting the story this way seems to make the students more attentive when we then begin to describe some of the key failures of classical physics that caused this dramatic shift in understanding.

The first type of experiment we discuss is atomic spectroscopy. We describe how atoms emit light only at specific frequencies. We then talk about Bohr's attempt to model this, saying that in his approach, the electrons orbit the nucleus only at certain radii, and that discrete light emission happens when electrons hop from one orbit to another. If the students are undergraduates, then this part of the lecture may include mathematical details since the Bohr model only requires the matching of classical forces and a few lines of elementary algebra. We normally skip these mathematical details for an audience of high school students. The students are then asked to explain what is qualitatively wrong with the Bohr model. In advanced placement physics classes, we usually find that a few students are able to identify the problem. At this point, we introduce the idea that electrons are standing waves that envelope the nucleus instead of point particles circling around it. The students are generally able to grasp the concept that these waves can be static and thus avoid the radiative instability issue that plagues the Bohr model. Of course, things like electrons still have particle-like properties due to their behavior in scattering processes, which is how they were first discovered. So the conclusion is that matter behaves like particles or waves depending on the situation.

\begin{figure}[h]
    \centering
    \includegraphics[width=0.4\columnwidth]{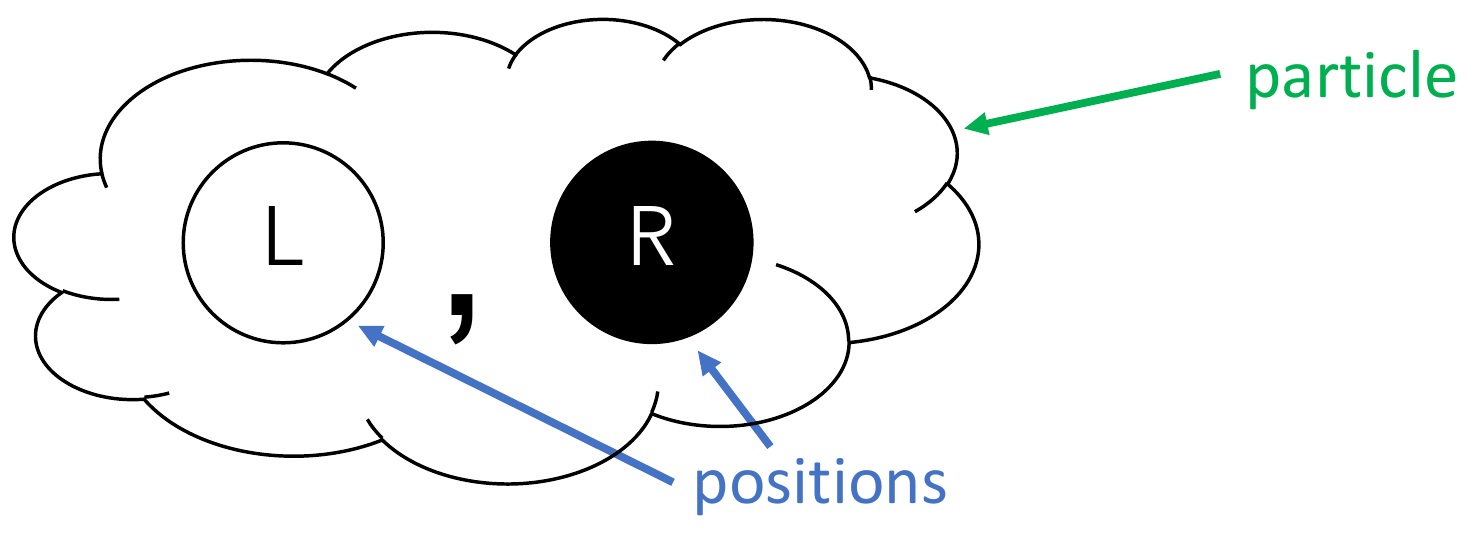}
    \caption{Depiction of superposition that ties the double-slit experiment to the friendly formalism described in the next section.}
    \label{fig:superposition}
\end{figure}

This brings us to the question: Waves of what? We use the double-slit experiment to answer this question. We play a six-minute excerpt from a video called ``Dr. Quantum" \cite{drquantumyoutube}. This video provides a very clear explanation of the experiment, including the effect of observations (note that the double-slit experiment is the only part we use---the movie it appeared in introduces a number of rather bizarre and misleading ideas about concepts from QM that should be avoided). Once the video concludes, we reiterate the part about what happens when electrons are injected one at a time and then explain how this leads to the conclusion that electrons (and all other ``particles") are waves of probability. These objects do not have definite locations until they are observed; until then, they are in a {\it superposition} of multiple positions. We use an image like that shown in Fig.~\ref{fig:superposition} to represent superposition. This image is designed to make a connection between the double-slit experiment and the formalism we use to describe classical/quantum information processing, as described in the next section. The fact that other properties like momentum and angular momentum are also probabilistic and subject to superposition is stressed as well. However, we mostly focus on position because this constitutes the most concrete and intuitive example for students at the high-school or early undergraduate level.

At this point in the lecture, the students invariably have a large number of questions, and considerable time is devoted to addressing these. We are often surprised at how astute the questions are; for example, high-school students (including one from CTech$^2$) have asked whether there might exist hidden properties that restore the determinism of Nature, essentially `inventing' the idea of hidden variables. Another high-school student asked what happens if measurements are repeated very quickly. We also spend time explaining that many of the conceptual questions are still unresolved to this day, including fundamental interpretations and what defines an observer. Schr\"odinger's cat is briefly discussed, as is the many-worlds interpretation (just to give a specific example of an interpretation that has been put forward). Spending at least some time on the open conceptual issues is especially important, because it lets students know that some of the confusion they may be experiencing is shared by the entire scientific community. We then stress that QM can still be used to make predictions and develop technologies even in the absence of a clear interpretation, and that the mathematical framework used to perform calculations is independent of any interpretation. We conclude the lecture by listing various existing technologies that rely on QM, including lasers, transistors, GPS, and magnetic resonance imaging.

This is the bare minimum that we do as an introduction to QM. If more time is available, then we may also discuss more experimental evidence for particle-wave duality, including the double-slit experiment for light, the photo-electric effect, Compton scattering, and blackbody radiation. We also discuss entanglement, the uncertainty principle, and spin if time permits.

\section{Overview of QIST lecture: A friendly formalism}\label{QI4Qformalism}

The QIST part of the lecture starts with some general background about the field, including quantum computing and quantum communications. We show the students some physical implementations of quantum bits that are currently pursued, as well as the state of the art in terms of the size of the devices and types of problems that can be run on them. The main part though is focused on introducing students to the QI4Q formalism. Normally, a challenge for students without a linear algebra background is that they do not have a means of doing meaningful hands-on activities and getting a sense of how quantum interference works. The QI4Q formalism allows them to perform calculations and discover on their own how circuits work. Below we describe this formalism. For more details, see Ref. \cite{Rudolph-book}. 

\begin{figure}[h]
    \centering
    \includegraphics[width=0.2\columnwidth]{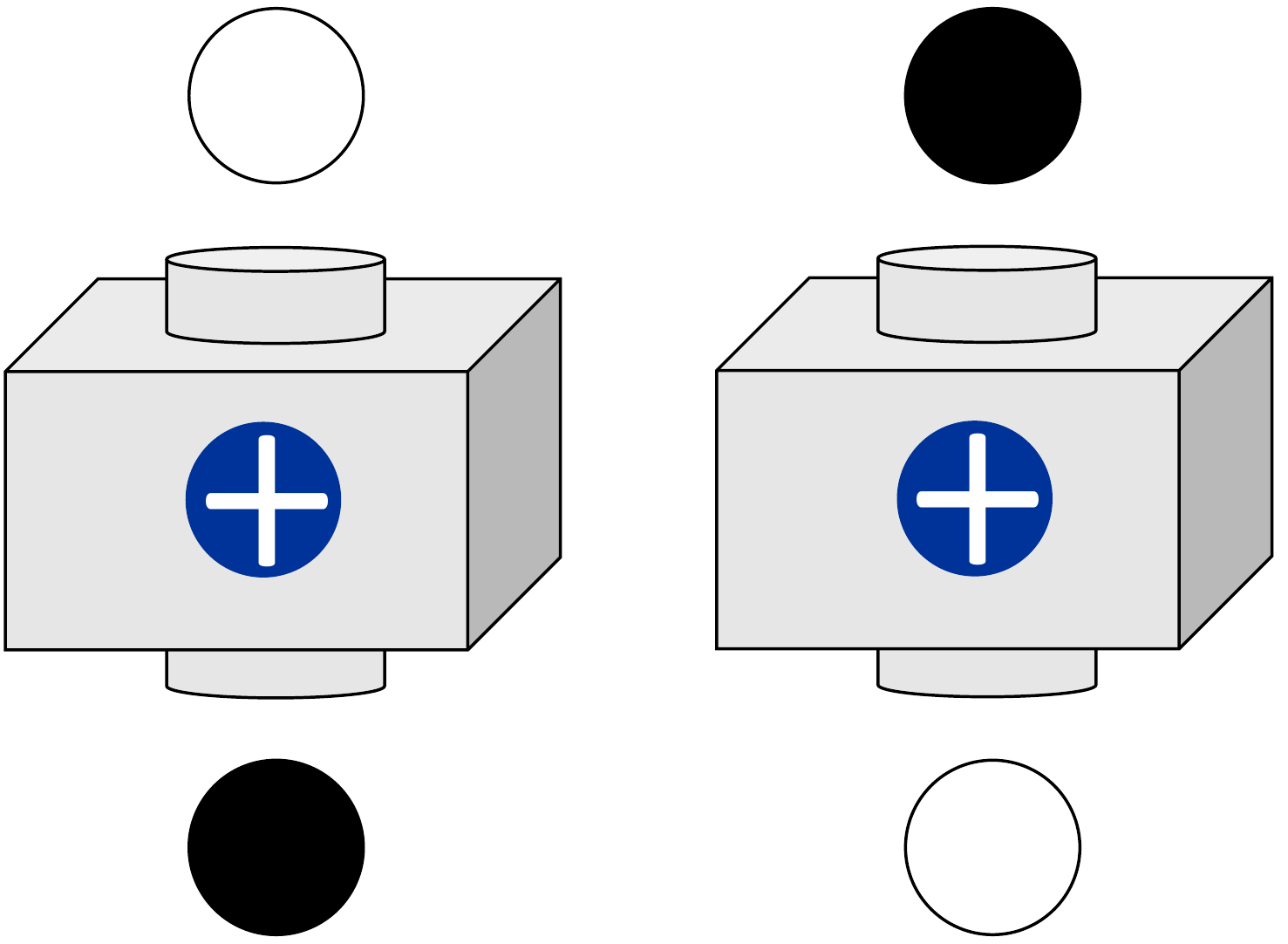}\hspace{2cm}
    \includegraphics[width=0.2\columnwidth]{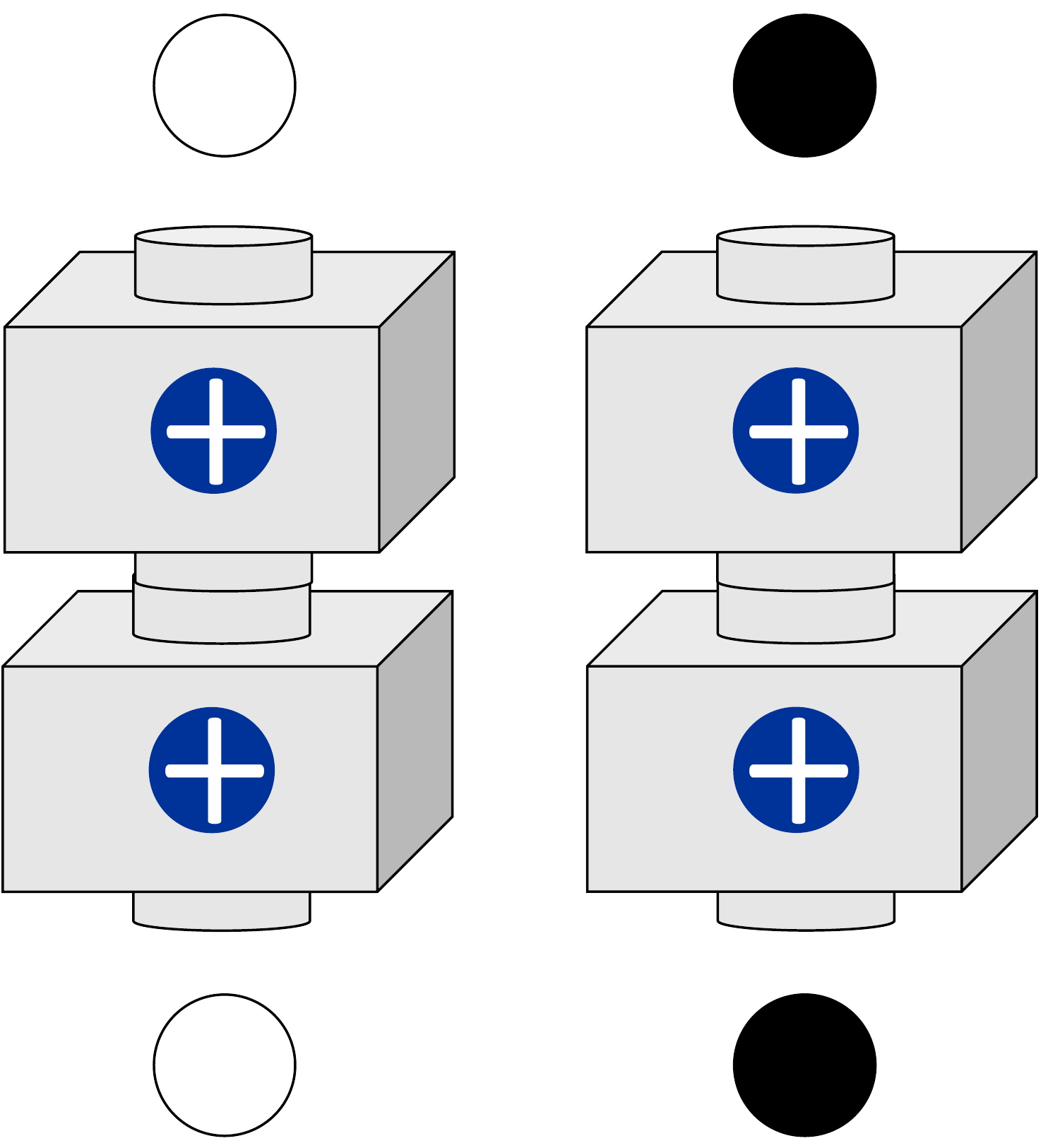}
    \caption{Basic properties of NOT gates.}
    \label{fig:NOTgate}
\end{figure}

\begin{figure}[h]
    \centering
    \includegraphics[width=0.65\columnwidth]{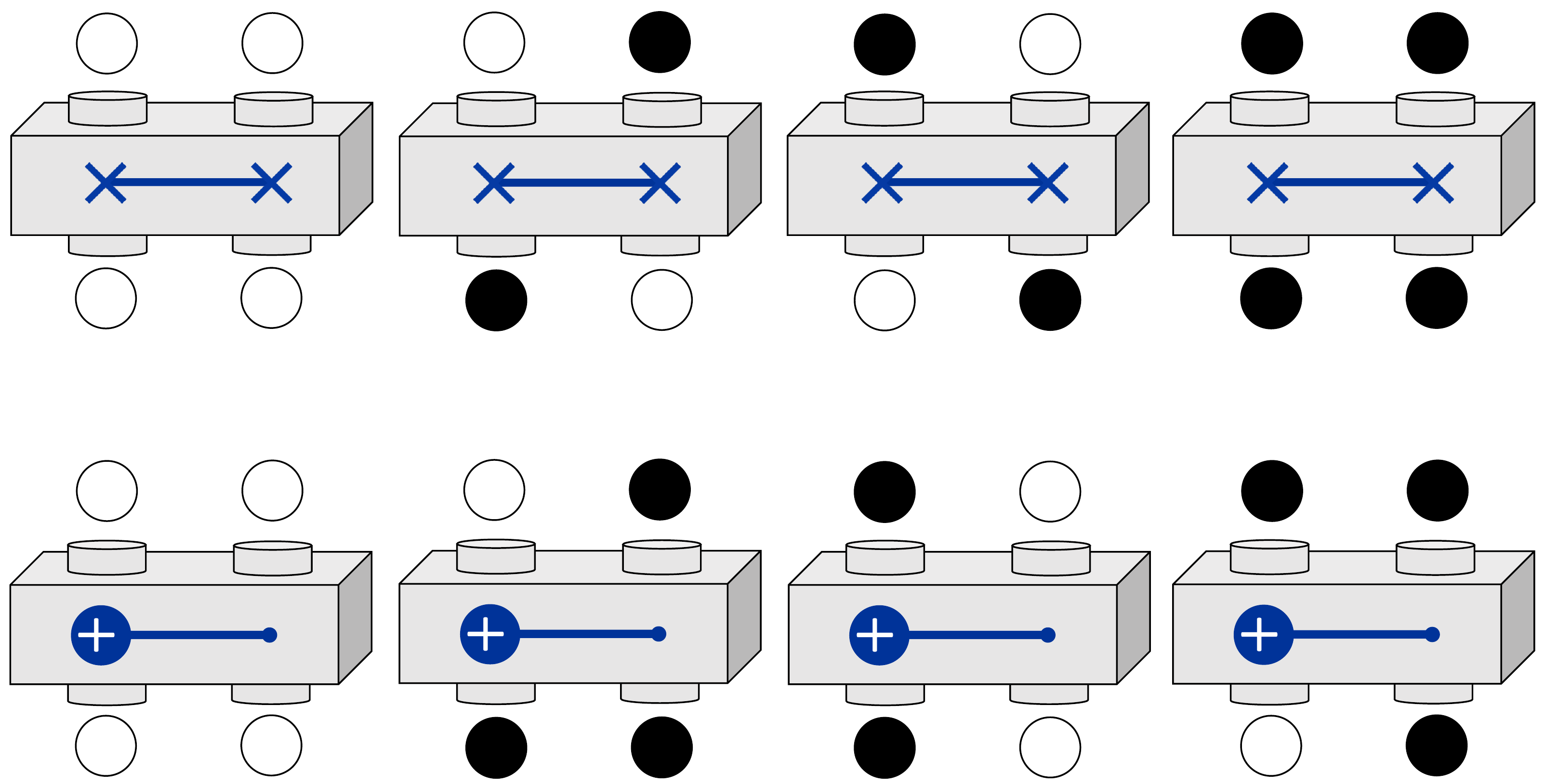}
    \caption{SWAP and CNOT gates.}
    \label{fig:SWAPandCNOTgates}
\end{figure}

The QI4Q formalism uses marbles to represent qubits, with white (black) representing state $|0\rangle$ ($|1\rangle$). Boxes that the marbles pass through represent quantum gates. Each box is labeled according to which gate it performs. We use the same notation and color coding as IBM Q experience instead of that of the QI4Q book, to make the connection easier for the students. Stacking boxes represents doing consecutive gates on qubits. We begin by introducing single-qubit and two-qubit gates that are also classically possible. First we present the NOT gate, which we label by X. The NOT gate switches the color of the marble, taking white to black and vice versa. Obviously, two X boxes stacked return the same color marble as what was input, as shown in Fig. \ref{fig:NOTgate}. Next we define two two-qubit gates: the SWAP gate, which simply swaps the colors of the two input marbles, and the control-NOT (CNOT) gate that applies NOT on one marble conditional on the color of the other, specifically only when the latter is black (the qubit is in state 1). The effects of both the SWAP and CNOT gates is shown pictorially in Fig. \ref{fig:SWAPandCNOTgates}. 

\begin{figure}[h]
    \centering
    \includegraphics[width=0.3\columnwidth]{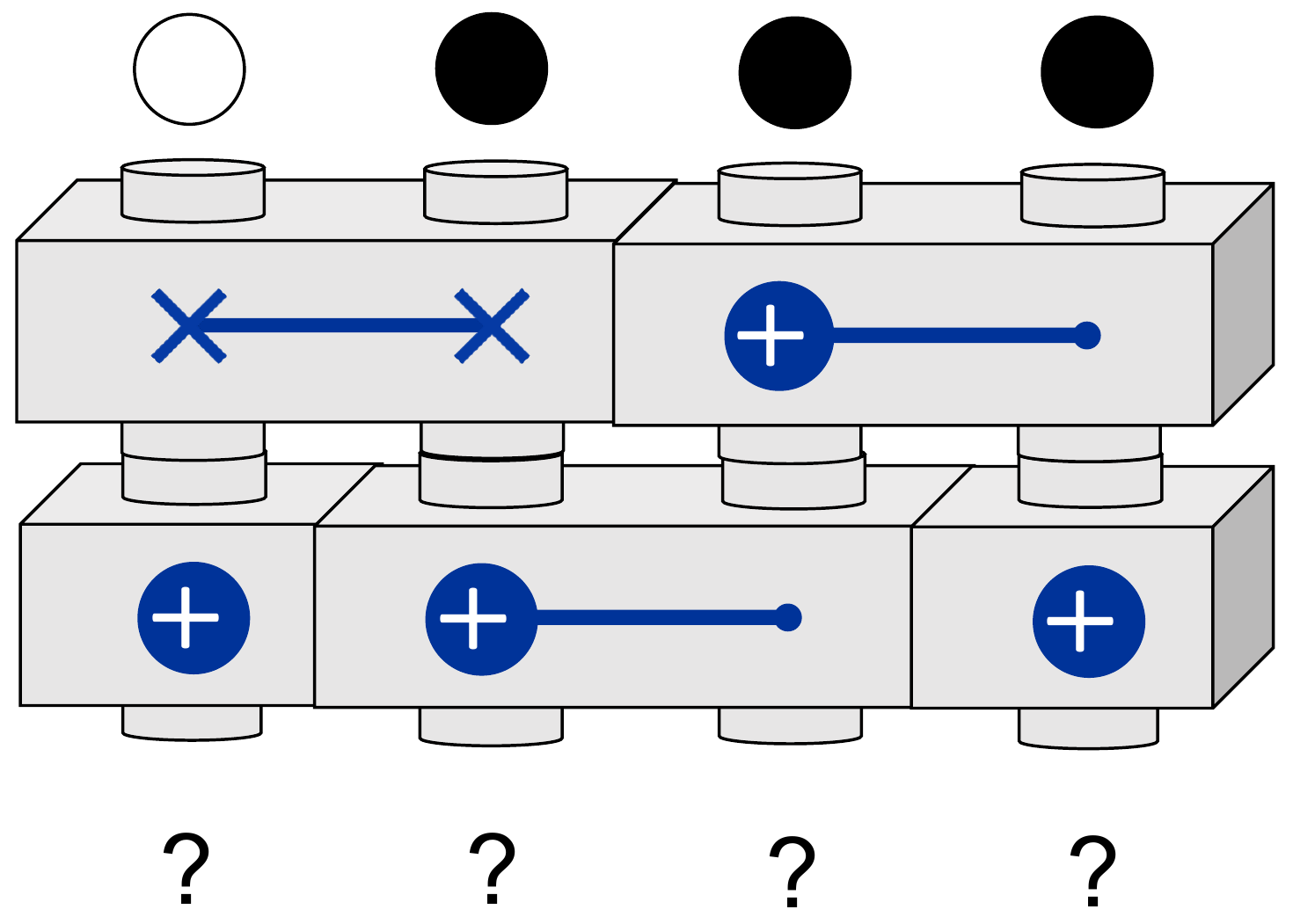}
    \caption{Exercise with SWAP and CNOT gates.}
    \label{fig:SWAPandCNOTgateExercise}
\end{figure}

At this stage, with the three `classical' gates introduced, the students are ready to run some simple (classical) circuits by stacking boxes and finding what the outputs are. The first such example they work on is shown in Fig. \ref{fig:SWAPandCNOTgateExercise}. This is the first hands-on problem the students are given, and typically they figure it out quickly. If more time is available, a few more examples of circuits made up of combinations of NOT, SWAP, and CNOT would be helpful in providing students with more practice. One could also consider asking students to come up with their own circuits. 

\begin{figure}[h]
    \centering
    \includegraphics[width=0.45\columnwidth]{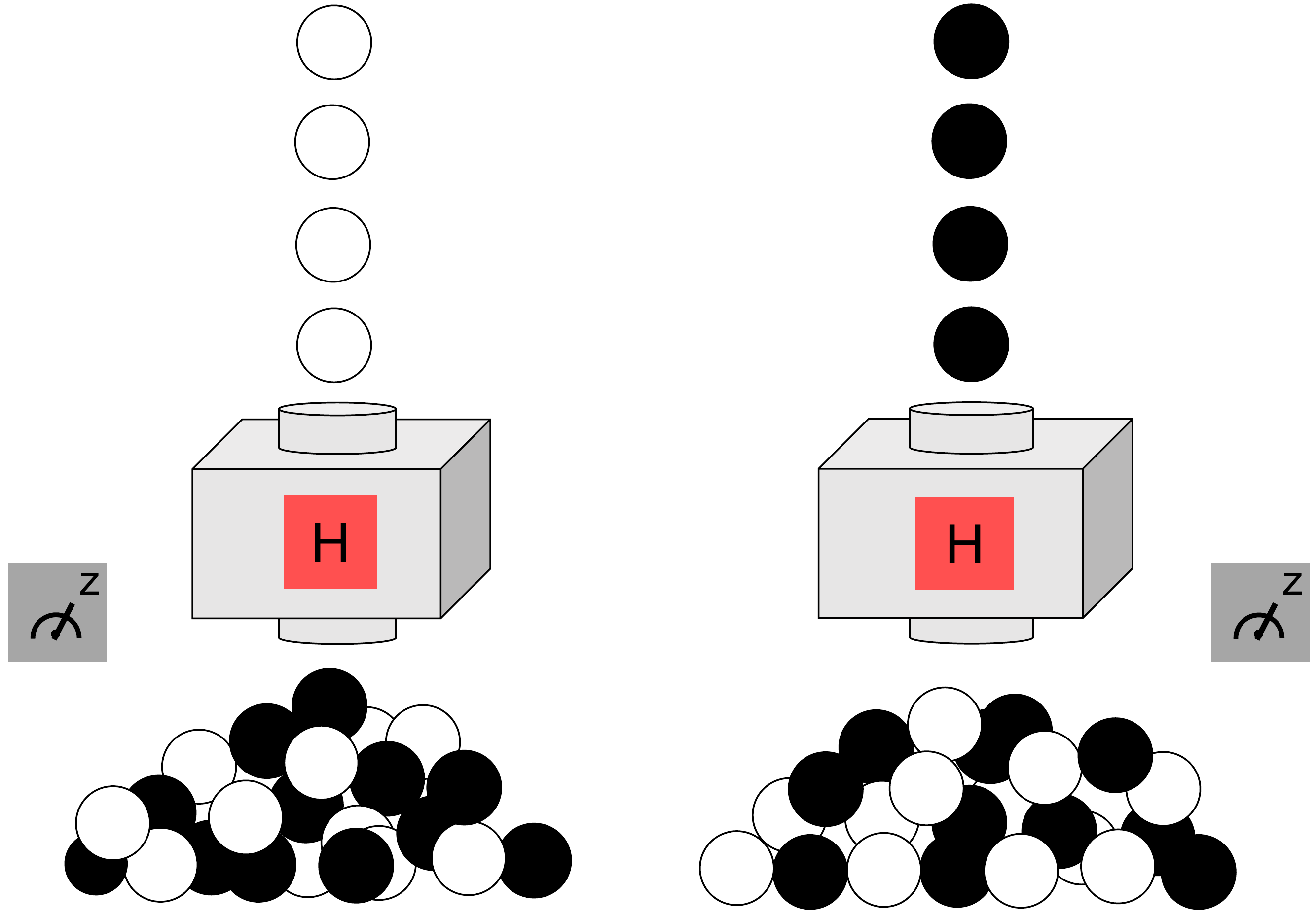}
    \caption{Introducing the Hadamard gate as a box that produces random outputs upon measurement.}
    \label{fig:Hadamardgate}
\end{figure}

\begin{figure}[h]
    \centering
    \includegraphics[width=0.2\columnwidth]{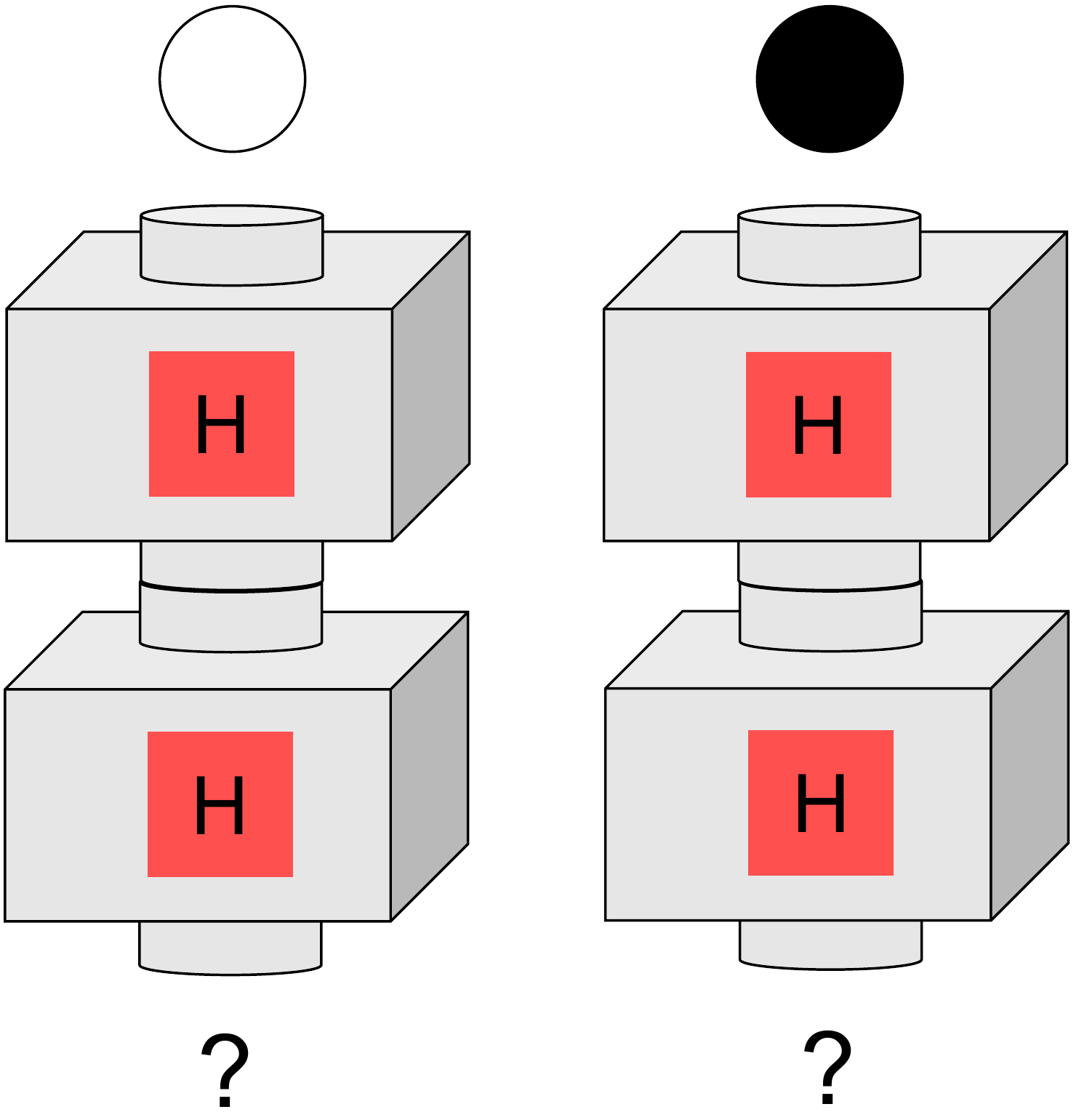}\hspace{2cm}
    \includegraphics[width=0.2\columnwidth]{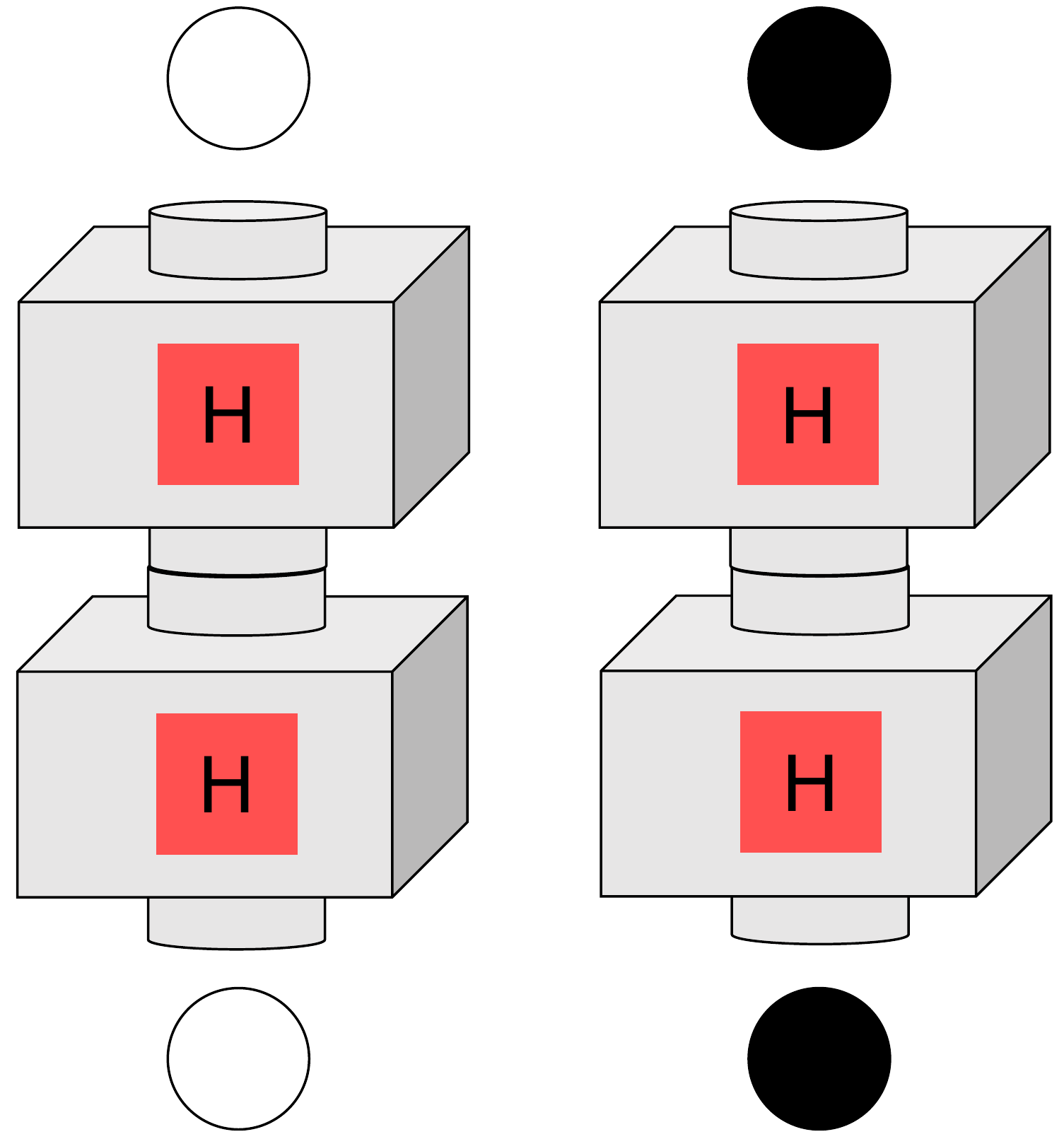}
    \caption{Showing that the random output of a Hadamard gate goes away when two of them are stacked together.}
    \label{fig:Hadamardgate2}
\end{figure}

The next step is to introduce a gate that is uniquely quantum, i.e., has no classical analogue: the Hadamard gate, labeled by H. This gate does something quite different: after passing marbles through it \emph{and measuring} the output color, we find that they are either black or white with probability 50\%, irrespective of the input color. Here it is critical to point out that a \emph{measurement} is made after the marble exits the H box. We use the IBM Q symbol to denote the measurement, as shown in Fig. \ref{fig:Hadamardgate}. At this point, we follow QI4Q and ask ``How can we interpret what the H gate does?" One may think that it is equivalent to flipping a coin and, depending on the outcome, setting the marble to either black or white. This is certainly consistent with what is shown in Fig. \ref{fig:Hadamardgate}. How can we check if this is indeed what is happening? What we can do is stack two H boxes and see what comes out. If we do this experiment, we find that the marble returns to its original color, as shown in Fig. \ref{fig:Hadamardgate2}, which of course contradicts the coin-tossing scenario, which would result again in a 50-50 probability of getting white or black at the output. 

\begin{figure}[h]
    \centering
    \includegraphics[width=0.3\columnwidth]{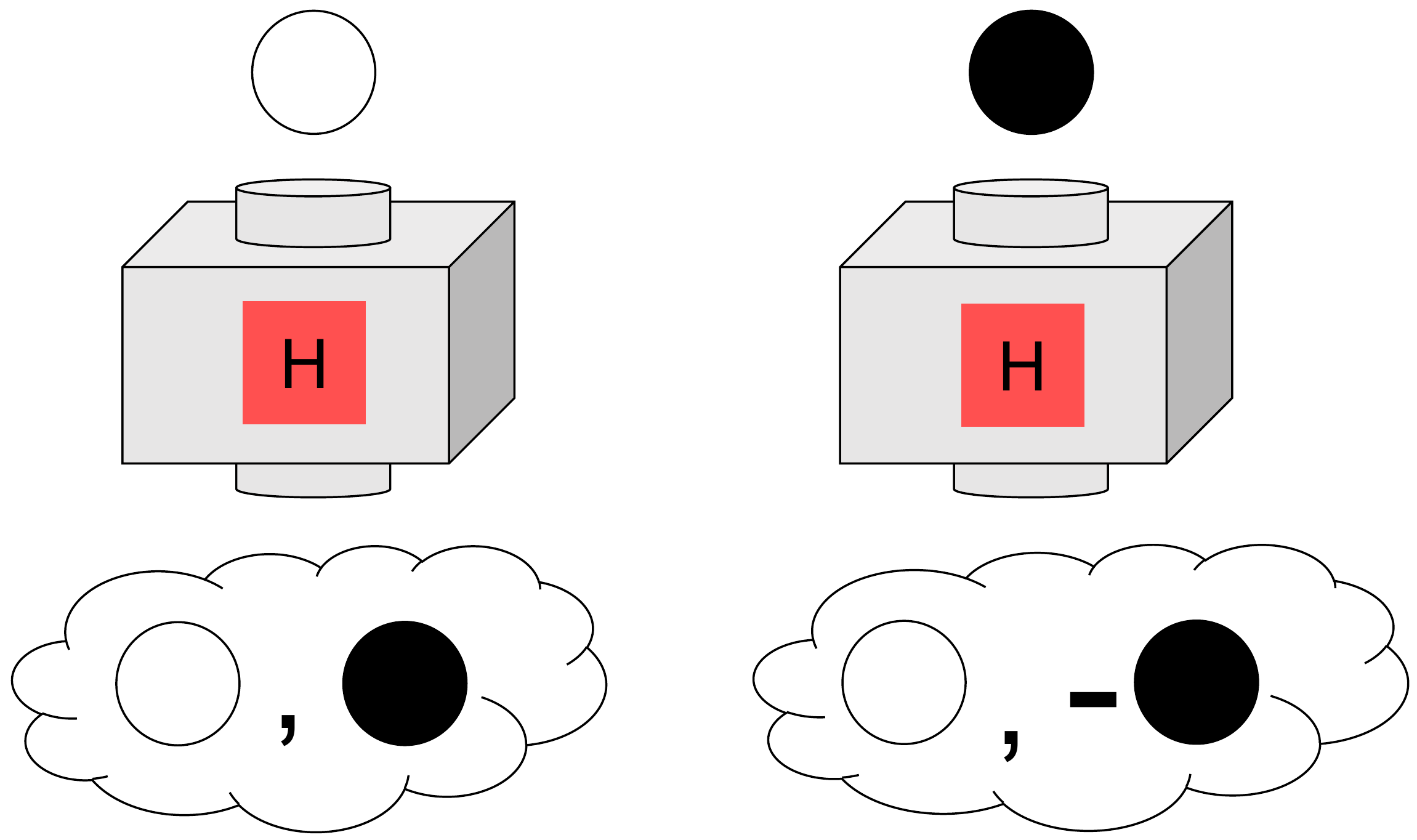}\hspace{1cm}
    \includegraphics[width=0.4\columnwidth]{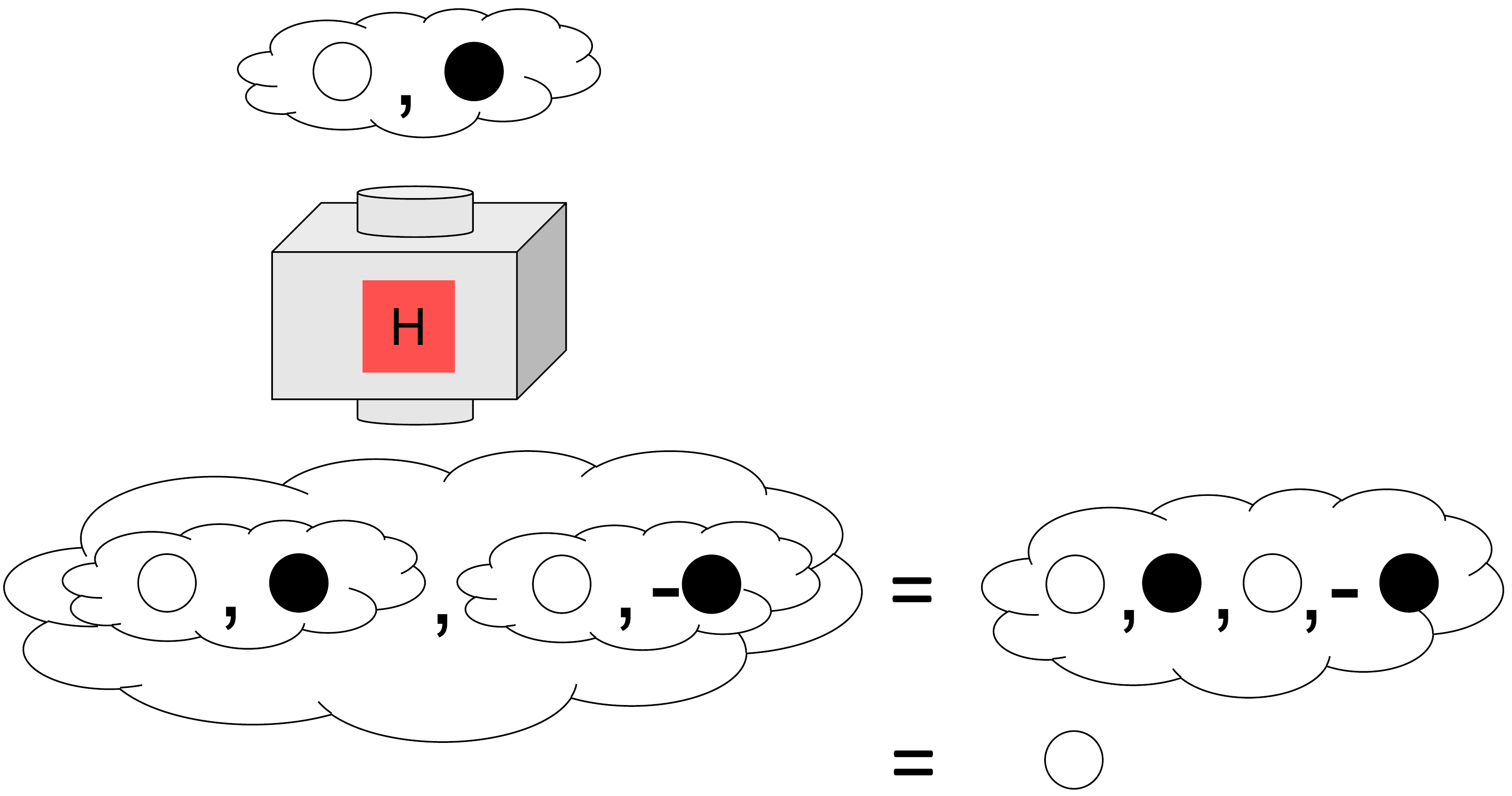}
    \caption{Hadamard gates produce misty states from non-misty ones and vice versa.}
    \label{fig:Hadamardgate3}
\end{figure}

It therefore becomes clear that we cannot represent the action of the Hadamard with our classical description of either black or white marbles, even if we introduce a probability associated with each color. We need something to represent the fact that the marble is in a superposition of white and black. At this point, we refer back to the interference of quantum particles discussed in the context of the double slit experiment and shown in the animated video. Continuing to follow QI4Q, we introduce the idea of a \emph{misty} state, which is another way to say a superposition. Such states are represented by clouds (`mists'), as shown in Fig. \ref{fig:Hadamardgate3}. Here we have to introduce in a somewhat \emph{ad hoc} manner a minus sign in front of the black marble in the output mist when the input marble is black. The appearance of this sign, which can seem rather mysterious to the students, is perhaps the most prominent spot where the deviation from a proper linear algebra treatment is apparent: in the case of linear algebra, the unitarity requirement would motivate the matrix entries, and it would be clear that the minus sign in front of the black marble allows the orthogonality requirement of the output states to be satisfied. Here, all that can be said is that the minus sign is used to distinguish certain misty states. Although measurements of the two output misty states on the left in Fig.~\ref{fig:Hadamardgate3} yield the same (random) results, there is a subtle difference between these two states that can be teased out if we combine the Hadamard box with other boxes. It also plays an important role if we input a misty state into the Hadamard box, as shown on the right side of Fig.~\ref{fig:Hadamardgate3} (the misty state manipulations here are explained in detail below). At any rate, the misty (QI4Q) formalism is very powerful, even if it comes with a few extra rules. The basic properties and rules for manipulating mists are listed below, where we also make use of a bracket notation to clarify some of the rules. For example, [W,B] denotes a superposition of white and black. Note that while we could use this notation alone (without the boxes, marbles, and mists) to describe quantum information processing, having a visual representation can render the material much more accessible to students~\cite{Kohnle2012,Kohnle2015}. 

\newpage

\subsection*{Rules for mists}
\vspace{-0.3cm}
\begin{enumerate}
\item A mist contains a series of entries separated by commas, where each entry is a collection of white and black marbles (one for each qubit), possibly with a minus sign in front. Each entry in a mist must contain the same number of marbles. This number is equal to the number of qubits. Here is an example of a mist with 3 entries describing 2 qubits: [WB,BB,BW].
\item The order of the different entries within a mist does not matter. For example, [WB,BB,BW] is the same as [BB,WB,BW]. However, the order of the marbles in a single entry {\it does} matter since each marble is associated with a different qubit. For example, [WB,BB,BW] is {\it not} the same as [BW,BB,BW].
\item If two entries contain exactly the same sequence of marbles but with opposite signs in front of them, then these entries cancel and can be removed from the mist. For example, consider the following mist: [WB,BB,BW,-BB]. This is the same as [WB,BW].
\item If a mist contains several identical entries (including same sign), then redundant entries can be deleted so long as the ratios of distinct entries remain the same. For example, [WW,WW,BB,BB] can be reduced to [WW,BB] because the ratio of WW to BB entries remains 1:1, but [WW,WW,WW,BB,BB] cannot be reduced because the 3:2 ratio would change. Note that this rule implies that if every entry in a mist is the same, then all but one entry can be deleted, e.g., [WB,WB,WB,WB] $\to$ [WB].
\item If there is only one entry in the mist, then the mist can be dropped, e.g., [WB] $\to$ WB.  
\item Mists within mists can be eliminated. 
\item If the first few qubits in every entry have the same sequence of marble colors, then these qubits can be factored out of the mist on the left. Similarly, if the last few qubits in every entry have the same color sequence, these qubits can be factored out to the right. This is similar to factoring out $a$ in an equation such as $ax+ay=b$ (making this analogy with the associative property of multiplication helped the high-school students in the C-Tech$^2$ program). 
\item Two or more mists can be combined into a single mist in a manner that is identical to the FOIL (`firsts, outers, inners, lasts') English-language mnemonic in algebra, e.g., $(a+b)(c+d)=ac+ad+bc+bd$.
\item Two mists related by an overall minus sign are equivalent, e.g., [-WB,-BB,BW]=-[WB,BB,-BW]=[WB,BB,-BW].
\end{enumerate}

\begin{figure}[H]
    \centering
    \includegraphics[width=0.62\columnwidth]{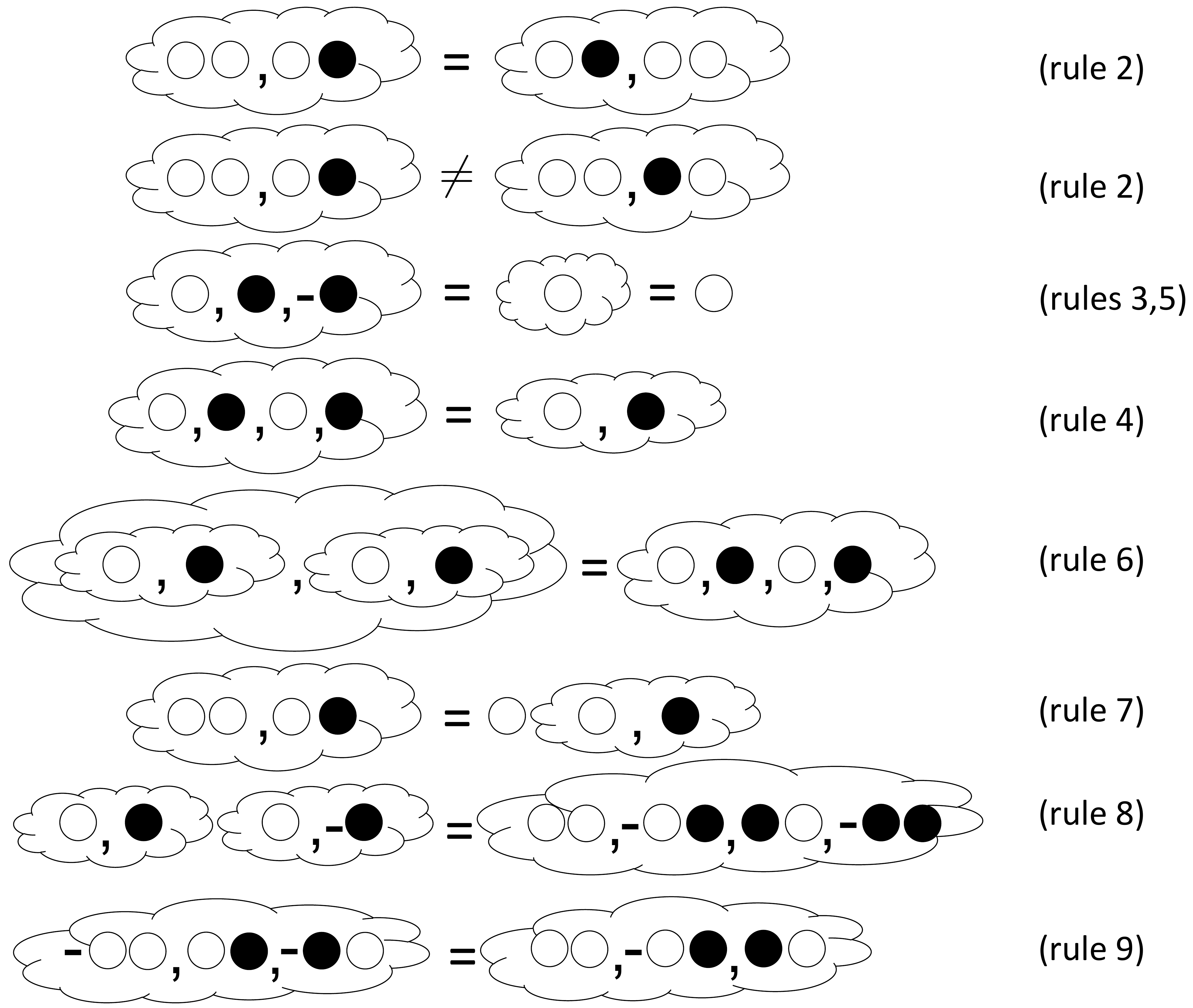}
    \caption{Rules for misty states.}
    \label{fig:cloudRules}
\end{figure}

A pictorial summary of these rules is given in Fig. \ref{fig:cloudRules}. Now that we have presented the mist rules, the next step is to explain how misty states pass through boxes. The basic rule for this can be summarized succinctly: Each entry in the mist is replaced by the output that results from passing that entry through the boxes. An example is shown in Fig.~\ref{fig:Hadamardgate3}, where the outer mist in the output represents the original mist from the input, while the inner mists result from passing each entry of the input through the box. Applying rule 6 from above, the output misty state turns into a mist with four entries, two white and two black, the latter with a minus sign difference between them. Applying rules 3 and 4, the black marbles cancel, and the white ones are replaced by a single white. Applying rule 5 then allows us to drop the remaining mist. This shows the phenomenon of interference that was discussed in the double slit experiment, which is here represented by the cancellation of the two black marbles. We see that in addition to producing misty states from non-misty ones, the Hadamard box can also turn misty states into non-misty states.

\begin{figure}[h]
    \centering
    \includegraphics[width=0.22\columnwidth]{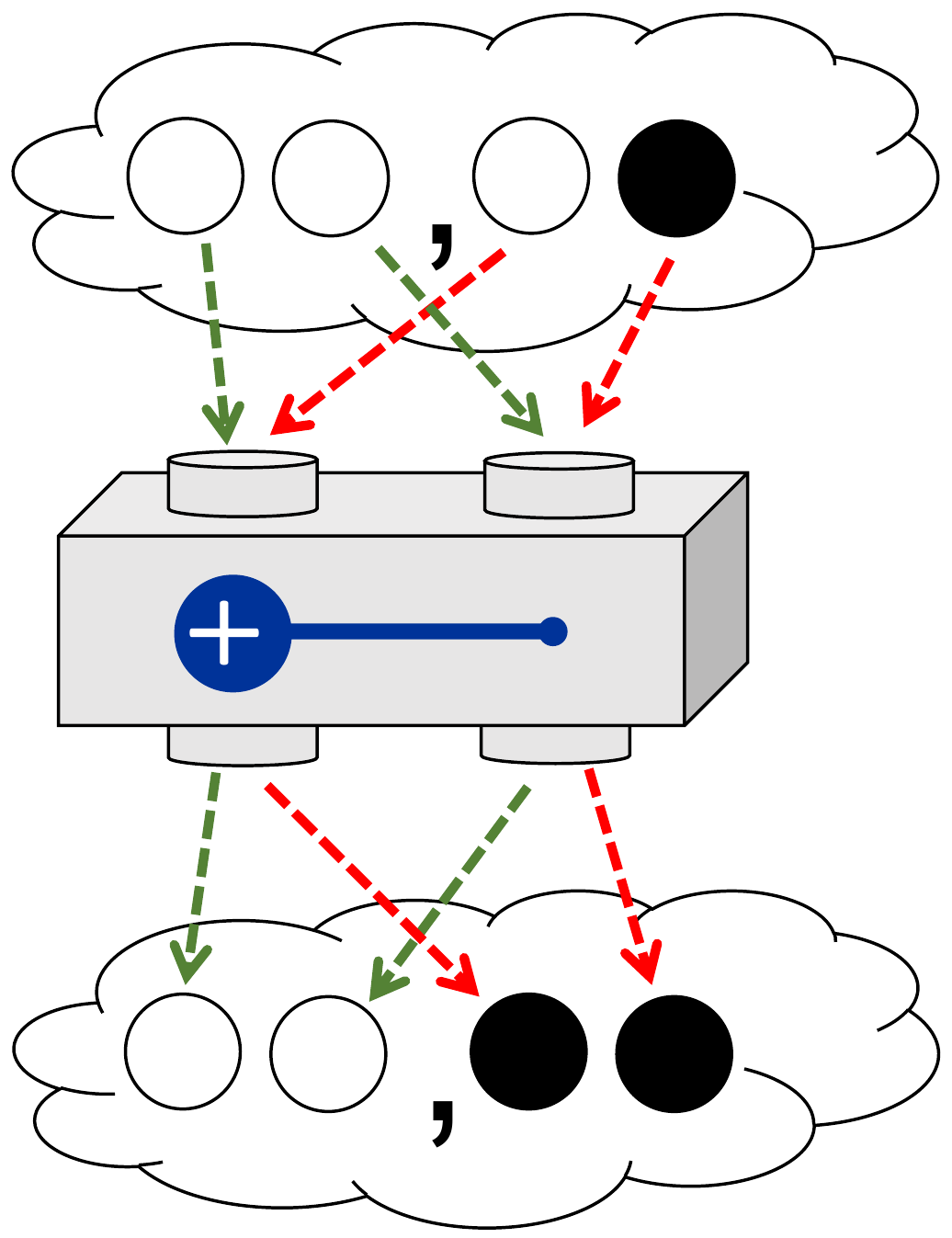}
    \caption{Inputting a two-qubit misty state into a CNOT box. In this example, entanglement is created.}
    \label{fig:entanglement}
\end{figure}

If a two-qubit misty state is input into a two-qubit box, the output can be obtained by separately passing each entry in the mist through the two-qubit box, just as we saw in the one-qubit case. An example involving the CNOT box is shown in Fig.~\ref{fig:entanglement}. The marbles in each entry should be inserted into the input ports in the same order that they appear in the entry, i.e., the left marble is inserted into the left port, the right marble into the right port. This generalizes straightforwardly to misty states with larger numbers of qubits. The fact that we pass each entry through the box(es) independently and collect the results together (separated by commas) to form the output misty state is a consequence of the \emph{linearity} of QM. Note that the example of Fig.~\ref{fig:entanglement} can be used to introduce the concept of entanglement since here we start with a factorizable misty state and obtain a misty state that cannot be factorized. With these ingredients, the students are ready to do hands-on exercises with misty states and boxes, which represent actual \emph{quantum circuits} in an unambiguous, mathematically precise way.

\section{IBM Q Experience}\label{ibmq}

The students are asked to create IBM Q Experience accounts before coming to the program. This is an important point, not only to save time, but also because of age restrictions IBM places on IBM Q users (as of the time of this submission, children under 14 years old cannot use the service, and those between 14-17 need a guardian to accept the agreement on their behalf).

Once the students have successfully accessed their accounts, some time is spent explaining the conventions. The symbols of the basic gates are familiar from the boxes introduced above. We next explain the circuit structure (which is standard in quantum computing) and its relation to the QI4Q boxes, along with the convention of which qubit in the box depiction corresponds to the top and bottom in the circuit representation. The students then create and run circuits on their own, starting with very basic examples and progressively going to more complex ones. The students go back and forth between working out examples by hand (using pen and paper or their tablets) and on the IBM Q site. This allows them to make predictions and then test to see that they get the expected results. We ask the students to start by using the simulator in IBM Q and later to also try running circuits on the hardware as well. This allows them to play with the simulators and see that the expected results are reproduced.  

\subsection{Simulator}

The students start with very simple circuits to verify the action of the gates they saw before. In particular, they created and run the following circuits (all followed by measurements of all the qubits):

\begin{enumerate}
\item X gate on one qubit
\item Hadamard gate on one qubit (here they  see the randomness in the outcome depicted in Fig. \ref{fig:Hadamardgate})
\item Two Hadamards on one qubit (here they  see how two Hadamards return the qubit back to its original state and that the result is deterministic)
\item Repeat, but apply a NOT gate before the Hadamards to see that the effect is the same for state 1 (black marble)
\item CNOT on each of the four input states WW, WB, BW, BB so that they verify agreement with what is depicted in Fig. \ref{fig:SWAPandCNOTgates}.
\item The circuit of Fig. \ref{fig:SWAPandCNOTgateExercise} to verify they obtain the same result as what they derived in the previous module
\item Hadamard on the first qubit follwed by a  CNOT with the first qubit as the control: this generates an entangled state. The students are asked to check that, while a measurement of one qubit only gives a random result, the outcomes when both qubits are measured are perfectly correlated; the term and concept of \emph{entanglement} are thus introduced. This is connected back to the QIST part of the lecture, where some of the technologies enabled by entanglement were discussed.
\end{enumerate}

\subsection{Hardware}

The students are asked to repeat a few simple exercises from above on the hardware. This allows them to appreciate that there are errors which give results that deviate from the error-free simulation. Indeed, one student who moved to the hardware on her own commented on the errors she was getting in the output (compared to the expected results and the simulator) and speculated that this is why we do not have quantum computers yet!

\section{Money or tiger game}\label{moneyortiger}

The formalism we introduced above allows students to get the main point of quantum algorithms without knowledge of linear algebra or even an introduction to the mathematical problem they are solving. We focus on a game developed by one of us (EB), which we call ``Money or Tiger". This is essentially Deutsch’s algorithm \cite{djalgorithm}, disguised as a game. Other papers have presented interesting and fun games that involve physical activities to help high-school students gain intuition about quantum concepts~\cite{LopezIncera2019,Svozil2006}. The game we introduce here is distinct from these in several ways. It does not require more than one student and relies on only pen and paper and the QI4Q formalism, and thus can be viewed as a preparatory step toward a proper linear-algebra treatment. Perhaps the most important distinction is that it introduces the concept of a quantum algorithm and the advantages that QM can bring to information processing. These are the two main learning objectives behind this game. It shows that a simple algorithm (combination of boxes) employing quantum gates can be used to solve a problem twice as fast as what can be done using only classical information processing.

\begin{figure}
    \centering
    \includegraphics[width=0.5\columnwidth]{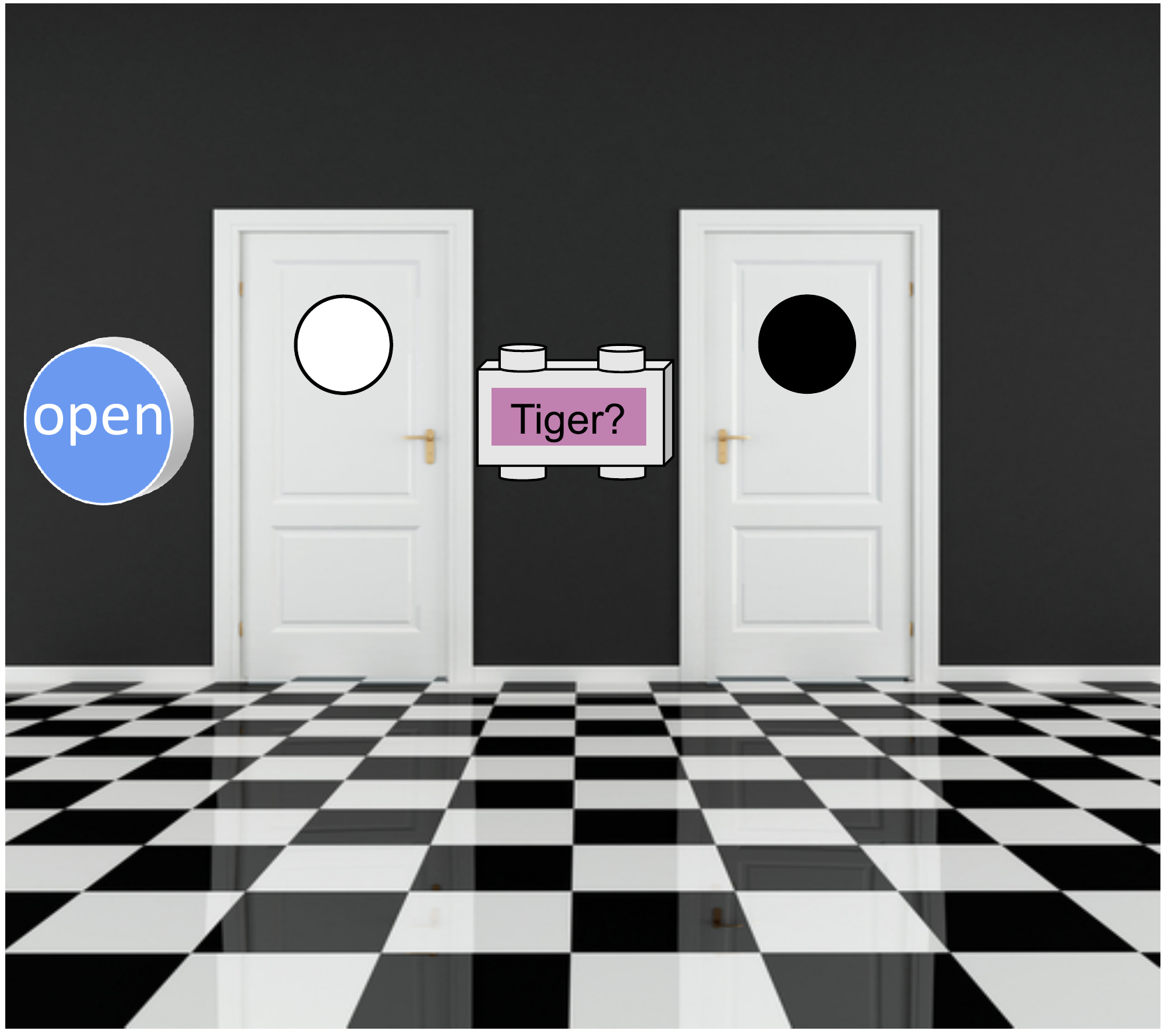}
    \caption{Setup of the Money or Tiger game.}
    \label{fig:setup}
\end{figure}

The setup of the game is shown in Fig.~\ref{fig:setup}. There are two doors, one labeled with a white circle, the other with a black circle. There is a button on the wall that opens both doors simultaneously. \textit{It is not possible to open only one door}. There is money behind at least one door. There may or may not be a tiger behind one of the doors. If there is no tiger, then you want to push the button and collect the money. However, if there is a tiger, then you do not want to push the button, and instead you leave without the money, happy enough that you are still alive.

\begin{figure}[h]
    \centering
    \includegraphics[width=0.4\columnwidth]{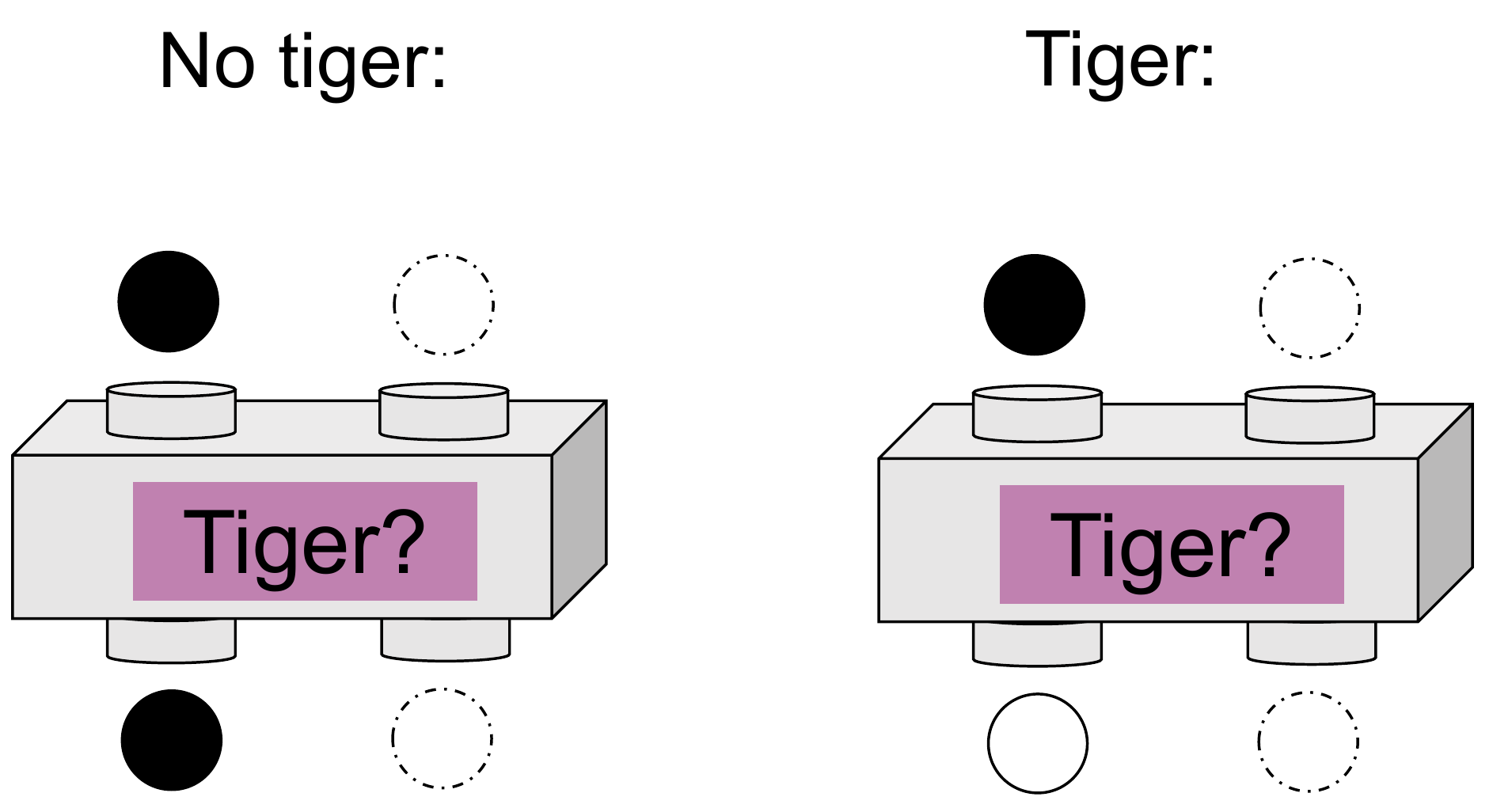}
    \caption{Tiger box rules. A black ``test" marble is always inserted in the left input port. In the right input port, insert a ``door" marble whose color corresponds to the door being checked. The door marble comes out the same color regardless of whether or not there is a tiger. However, the test marble changes color if a tiger is present.}
    \label{fig:tigerBoxRules}
\end{figure}

Also on the wall is a box labeled ``Tiger?". You are allowed to query this box once (and only once) to check whether there is a tiger. The way the box works is as follows. The box has two input ports and two output ports. You always input a black marble in the left input, and in the right you insert a marble whose color matches the door you want to check. If you want to know whether there is a tiger behind the white door, then you insert a white marble, while to check if there's a tiger behind the black door, you insert a black marble in the right input port. The marble that emerges from the right output port is always the same color as what was inserted into the right input port. However, the color of the marble that comes out of the left output depends on whether or not there is a tiger behind the door being checked. If there is no tiger, this marble is black, while if there is a tiger, it is white. These rules are summarized in Fig.~\ref{fig:tigerBoxRules}. 

If we only have access to classical information processing, then it is clear that the Tiger box needs to be queried twice in order to be sure there is no tiger present. You would have to use it once for each of the two doors. The point of this game is to show that QM allows us to determine whether or not there is a tiger behind either one of the doors with 100\% certainty while only using the Tiger box once.

\begin{figure}
    \centering
    \includegraphics[width=0.95\columnwidth]{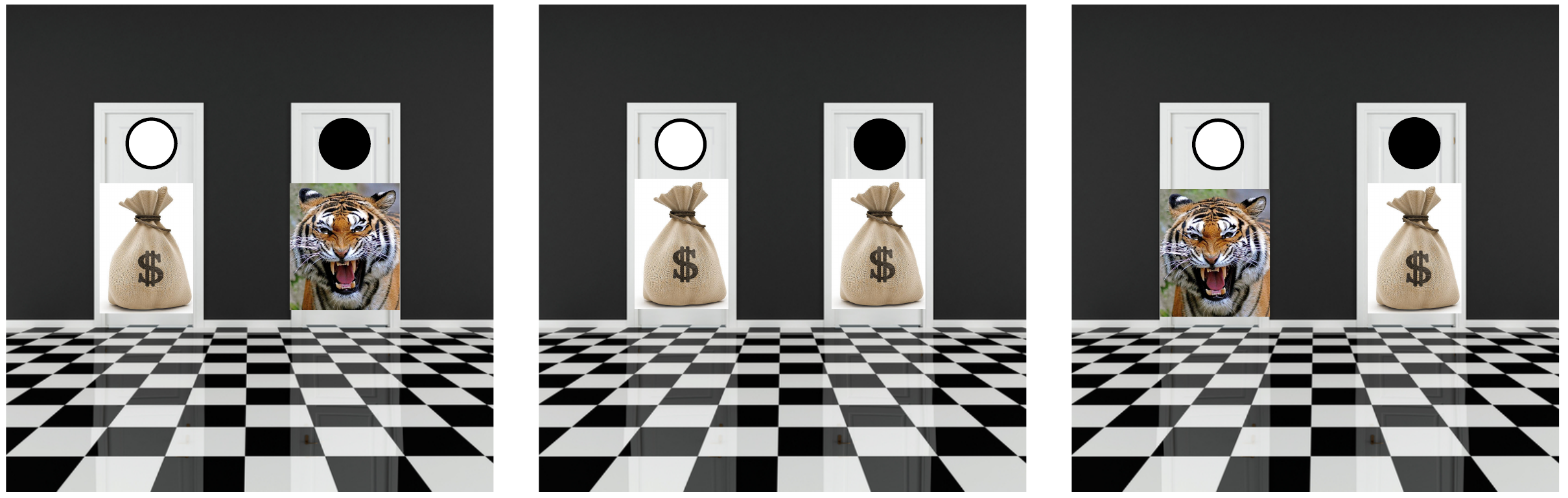}
    \caption{Three possibilities for the Money or Tiger game. There is money behind at least one door and either (left) a tiger behind the black door, (middle) no tiger anywhere, or (right) a tiger behind the white door.}
    \label{fig:3possibilities}
\end{figure}

After explaining the rules of the game to the students, we then break down the game into two steps. We first point out that there are three possibilities for what is behind the doors. These are shown in Fig.~\ref{fig:3possibilities}. The students are asked to find circuits that have the same rules as the tiger box in each case. To provide further guidance, they are told that they should only use NOT and CNOT gates. The solution is: If there is a tiger behind the black door, then the Tiger box is a CNOT. If there is a tiger behind the white door, the Tiger box is a CNOT with a NOT gate added below the left output port. If there is no tiger, the Tiger box is an identity gate. The problem then becomes how to determine which of these three circuits the Tiger box corresponds to.

\begin{figure}[h]
    \centering
    \includegraphics[width=0.15\columnwidth]{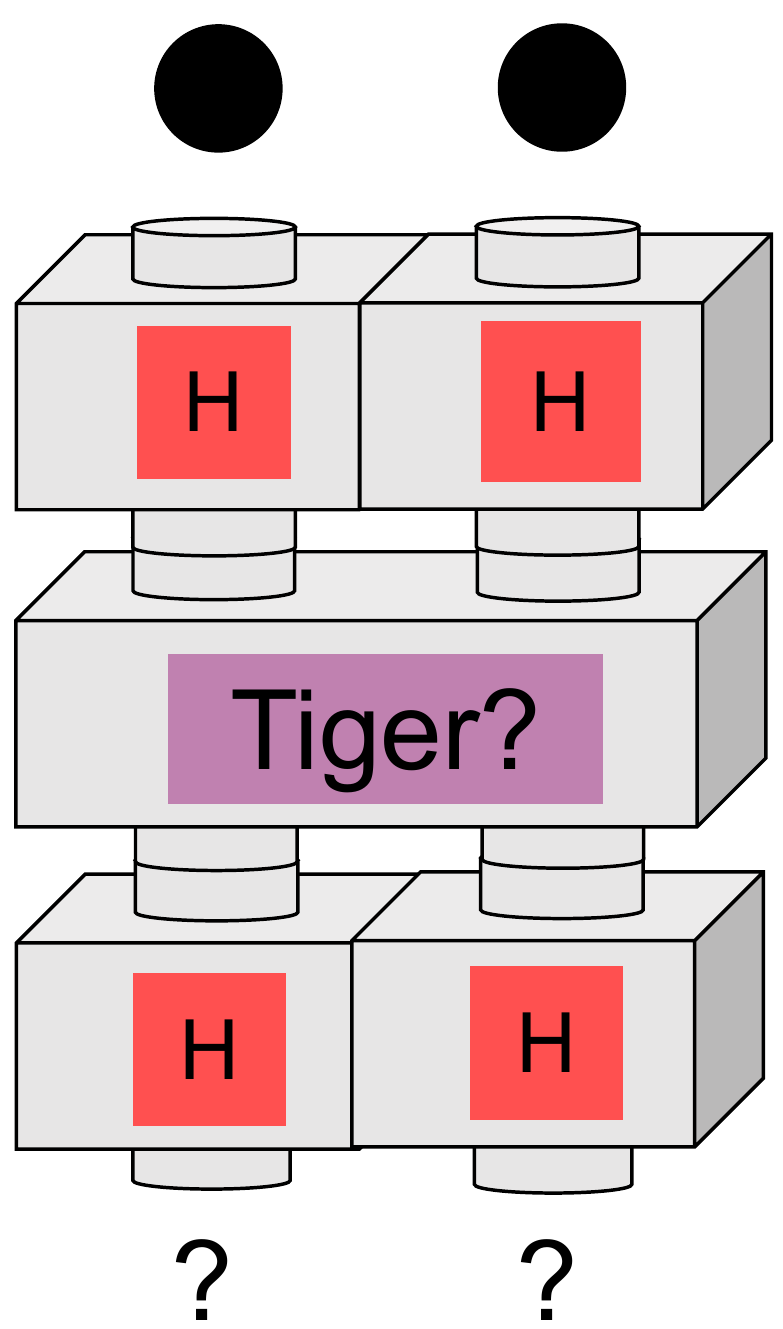}
    \caption{Circuit that solves the Money or Tiger game.}
    \label{fig:moneyOrTigerSolution}
\end{figure}

\begin{figure}
    \centering
    \includegraphics[width=0.4\columnwidth]{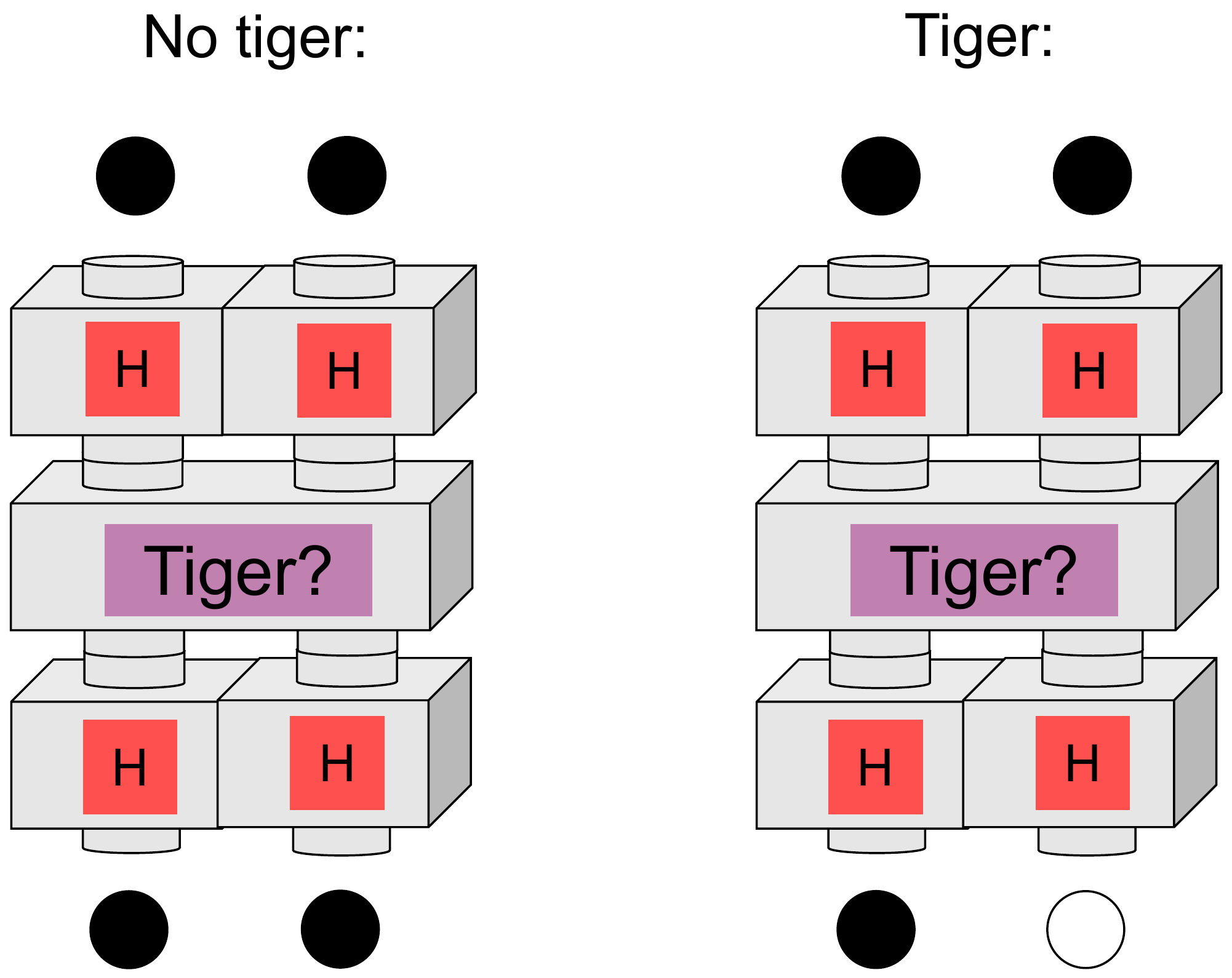}
    \caption{Outputs of the solution circuit when there is or is not a tiger. If two black marbles are input into the circuit, then a white output signifies the presence of a tiger, regardless of which door the tiger is behind.}
    \label{fig:moneyOrTigerSolution2}
\end{figure}

The next step is to show that by adding additional boxes above and below the Tiger box, it is possible to determine whether or not a tiger is present in one shot. Depending on how much time there is, we either ask the students to try to figure this out themselves, or we show them the answer and ask them to verify it. In the former case, they are told to combine Hadamard boxes with the Tiger box. In the latter case, we ask them to work out what happens when two black marbles are inserted in the circuit shown in Fig.~\ref{fig:moneyOrTigerSolution} for all three cases depicted in Fig.~\ref{fig:3possibilities} both by hand and by using the IBM Q hardware. The answer is shown in Fig.~\ref{fig:moneyOrTigerSolution2}, where we see that if there is no tiger, both outputs are black, while if there is a tiger, one output is white, regardless of where the tiger is.

The students see that unlike the classical case, where the box needs to be used twice, in the quantum case a single use of the tiger box suffices to identify the presence of a tiger.  Note that if the box is used twice in the classical setting, we also find out which door the tiger is behind. In the quantum case, where the tiger box is only used once, we only found whether there is a tiger, but not which door it is behind. This is analogous to the Deutsch algorithm, where we find out using the quantum circuit whether a function is balanced or constant, but not which particular function it is. In Ref.~\cite{Rudolph-book}, a larger variation of the Money or Tiger game shows that it is possible for the quantum case to only require a single use of a box when the classical case requires a large number of uses. This helps a student appreciate that the distinction between quantum and classical computing is about the number of algorithmic steps, and not about smaller and faster hardware or other similar misconceptions.

\section{Additional features and limitations of the QI4Q formalism}\label{limitations}

The rules (1) - (8) discussed above are not completely exhaustive. The two omissions (unnecessary for the introductory examples we have discussed) are (i) the rule for computing the probability of seeing some particular configuration of marble colors given a generic misty state and (ii) the rule for computing how the misty state changes (wavefunction collapse) if only one marble of a multi-marble state is observed. Given these two additional rules we can, in principle, use the QI4Q formalism to reproduce all the key results of quantum information such as teleportation, Bell’s theorem, no-cloning theorem, Shor’s algorithm, dense-coding, etc. This is because the formalism can deal with gates that are universal for quantum computing (although the set of gates described above are not actually universal - one has to include a nontrivial three-marble gate as well, which is easily done). 

While the QI4Q approach is (presumably!) an impractical way to teach all of modern quantum theory, the fact that it is possible in principle may reassure the instructor that listing the seemingly arbitrary rules for marble mist manipulation is no worse than the many classic undergraduate quantum theory textbooks which begin by listing the mathematical postulates of quantum theory (phrased, of course, in terms of linear algebra). 

It is worth noting several mathematical and conceptual differences between the abstract formalisms - an instructor who casually mixes up concepts between them can cause considerable confusion. The QI4Q formalism is comprised of a series of \emph{string rewriting} rules. That is, we start with a ``well formed'' string like [WB,BB,-WW,-WW] and then, depending on the operational scenario we are confronted by, we apply some string manipulations that are guaranteed to generate another valid string. Finally, there is necessarily a bridging rule/principle for mapping between the mathematical object and our operational interventions with, and observations of, the physical world.

Often, as seen in some of the examples above, the rules that generate valid strings from previous valid ones resemble the rules for manipulating numerical expressions involving addition and multiplication. It is important to not make too much of this; the underlying mathematical structures are ultimately very different, and a student who thinks of the correspondence as more than a helpful mnemonic can end up with nonsense: $2\times 3 =3\times2$ and $2+3=3+2$ but $\mathrm{[WB]}=\mathrm{[BW]}$ is an incorrect string equivalency, while $\mathrm{[W,B]}=\mathrm{[B,W]}$ is correct.

It is tempting to try and gradually sneak in some linear algebra while using the QI4Q rules. For example, one may think it suitable to introduce vectors that simply list the number of times a particular configuration of marble colors appears:
\begin{equation}
    \mathrm{[WW,WB,BB,WW]} \leftrightarrow 
    \left(\begin{matrix}
    2 \cr
    1 \cr
    0 \cr
    1
    \end{matrix}\right).
\end{equation}
Perhaps one could also introduce normalization to simplify the computation of probabilities, matrices to simplify the process of determining what transformation a given box implements, and so on. Are these really simplifications? There is considerable overhead associated with linear algebra. For instance, the students must now rigorously keep the correct ordering of entries, and new ordering rules need to be followed once additional marbles enter the picture. This is unnatural compared to understanding a superposition as an unordered set or list containing all the various physical alternatives that can potentially be observed, which accords with the use of lists from their everyday experience. Similarly, the tensor product is an advanced piece of linear algebra. But the Cartesian product (which is what the QI4Q approach uses in its place) is a completely implicit and standard part of any person's everyday thinking  about combining alternatives for disparate phenomena.

There is one feature that axioms stated in terms of linear algebra can readily capture which is not trivial in the QI4Q approach, and that is a way of specifying what are the allowed gates (boxes) within nature? In terms of linear algebra we simply state that any unitary transformation of the quantum state vector is (in principle) realizable by some physical setup. But in the QI4Q formalism as described above, the only non-classical gate is the Hadamard. A student may well ask (and, in fact, has!) whether it is possible to do a `controlled-Hadmard'. Na\"ively, it seems that this should be possible, since a `controlled-Hadamard' is certainly a valid quantum gate/unitary transformation.

However, the controlled-Hadamard is not implementable on marbles in misty states. It would be natural to try and define it as `a box which does the Hadamard on the first marble if and only if the second marble is black', in direct analogy to the CNOT, which does an X on the first marble if and only if the second (control) marble is black. Applying the controlled-Hadamard operation to the misty state [WW,WB,BB] leads to [WW,[W,B]B,[W,-B]B] = [WW,WB,WB]. A student who has learned the extra rules alluded to above regarding computing probabilities would incorrectly deduce that the probability to find both marbles being white if we made a measurement is $(1^2)/(2^2+1^2)=1/5$. This does not match the correct prediction for sending two qubits initially in the state $\ket{00}+\ket{01}+\ket{11})/\sqrt{3}$ through a controlled-Hadamard gate. 

What has gone wrong is that it is only a subset of quantum gates that can be rigorously mapped into the QI4Q picture, namely the ones that are given by a unitary matrix that is \textit{proportional to} a matrix of integer values. The constant of proportionality can be irrational (as in the Hadamard gate). For the controlled-Hadamard there is no constant it can be multiplied by to leave integer entries. Obviously there are many other quantum gates that the QI4Q formalism does not automatically capture (all the ones using complex numbers for example!) but it is a bit disconcerting that one with a seemingly simple pedagogical description will not work. 

This realization highlights the somewhat remarkable fact that universal quantum computing is possible using only quantum gates whose entries are proportional to integers \cite{adleman-qcomputability}. From that result we know that with only a small overhead in the number of marbles used, we can simulate the evolution of \emph{any} quantum gate: controlled-Hadamards, complex valued gates, and, in fact, any quantum systems whatsoever. The rules for QI4Q, which seem so much simpler than those of ``full quantum theory'', in some sense have just as much power.

The discussion so far has focused on the pragmatic calculational equivalence of the formalisms: For the operational setups that both approaches can deal with, they will agree on any possible observable statistics, and for ones the QI4Q approach cannot deal with automatically, it will incur only a limited overhead. But it is important to emphasize that what is going on is not an isomorphism. This can lead to some interesting conceptual/foundational points, ones that may be of interest to advanced students. 

For example, a white marble that passes through two Hadamard boxes emerges in the mist [W,W], and after passing through four boxes it emerges in [W,W,W,W]. Operationally, we obviously only see the marble to be white. The mist has, however, implicitly retained a count of the number of Feynman paths (using the quantum circuits definition of such) that the linear algebra would not reveal (since the output state would always be $\ket{0}$). The mist contains a little more of a record of its history than the quantum state (which is one of several reasons for using distinct language for it).

While this seems somewhat trivial (it is due to the fact we are using something equivalent to unnormalized quantum state vectors) and easily rectifiable by introducing the appropriate simplification rule within the QI4Q approach, from the viewpoint of quantum foundations it can be an interesting launching point for a discussion along the following lines: The mist and/or quantum state is a mathematical object that we humans find computationally useful, but its correspondence to ``the things really going on'' is contentious. Would an \"uber-powerful being that can see the real state of a physical system be able to deduce anything at all about how it arrived in that state or not?

The answer is ``no'' if the standard linear-algebraic quantum state is the real state (because even if the initial state is known there are infinitely many unitary evolutions connecting it to the final state that the \"uber-being can see). 

The answer is ``the \"uber-being would know the total unitary a system had experienced in its lifetime'' if the real physical state that they can observe is given by Heisenberg-picture operators (see, e.g., the Deutsch-Hayden \cite{DeutschPRSLA2000} construction). That is, even within standard quantum theory, operationally equivalent approaches can have different foundational implications. Note that knowledge of the Heisenberg operators would still not let our \"uber-being know the specific sequence of interactions (the quantum circuit) that implemented that total unitary, but it is, in fact, possible to construct other operationally equivalent approaches that do contain such information.

The upshot is that the differences between the standard and QI4Q formalisms can be a learning experience, as long as careful distinction is always made between the mathematics and the observed physical phenomena. Trying to gradually morph the one method into the other creates a potential for confusion. The full machinery of linear algebra is critical when (and only when) the situation being considered is complicated enough that there are geometrical constructions and intuitions to simplify a calculation. A student that gets to the point of analysing such complicated scenarios is presumably adept enough mathematically to translate between the two formalisms.

\section{Conclusions}

In conclusion, we have presented an accessible outreach program for high-school and early undergraduate students that does not require any advanced mathematics, and which is flexible and can be used for a quick exposure of students to the ideas of QIST (e.g., over a few hours or days) or expanded to a short course lasting several weeks. Our program has four main components: a short lecture on QM and QIST, a hands-on session for learning the QI4Q formalism, a hands-on session for practicing with the IBM Quantum Experience, and a hands-on session in which students either solve or verify the solution of the Money or Tiger game, which is essentially Deutsch's algorithm. We believe that our program can be used as an outreach tool both by QI4Q specialists and by high-school teachers. We hope to inspire others to use our tools and build on this program.

\section{Acknowledgements}

EB and SEE acknowledge support from an NSF EFRI program (Grant No. 1741656). EB also acknowledges support from NSF Grant No. 1847078.

\appendix

\section{Learning goals}

The learning goals for each part of the program are as follows. For the QM lecture, the goal is for students to learn the basic concepts of QM, especially superposition and probability. The historical approach we take is meant to facilitate this process; showing what motivated scientists to develop QM can help students better absorb the basic ideas and appreciate the conceptual challenges that still persist. This lecture provides them with the basic background needed to understand the main ingredients used in the subsequent QIST lecture. The goal of the QIST lecture is for the students to learn that there is a range of possible technological applications and intense ongoing efforts worldwide to realize them. The first learning goal of the part of the program focusing on the QI4Q formalism is for the students to understand that there exist quantum gates which do not exist classically and which can create new types of states. The second goal is that the students learn how to use the QI4Q formalism to analyze and design quantum circuits. The learning goals of the section focusing on the IBM Q Experience are first for the students to understand the mapping between the marble/box formalism and the quantum circuit and second for them to learn how to use the IBM Q Experience. Finally, the objectives of the Money or Tiger game include learning what a quantum algorithm is and understanding how certain tasks can be performed faster using quantum computers compared to classical computers.

%


\begin{thebibliography}{21}%
\makeatletter
\providecommand \@ifxundefined [1]{%
 \@ifx{#1\undefined}
}%
\providecommand \@ifnum [1]{%
 \ifnum #1\expandafter \@firstoftwo
 \else \expandafter \@secondoftwo
 \fi
}%
\providecommand \@ifx [1]{%
 \ifx #1\expandafter \@firstoftwo
 \else \expandafter \@secondoftwo
 \fi
}%
\providecommand \natexlab [1]{#1}%
\providecommand \enquote  [1]{``#1''}%
\providecommand \bibnamefont  [1]{#1}%
\providecommand \bibfnamefont [1]{#1}%
\providecommand \citenamefont [1]{#1}%
\providecommand \href@noop [0]{\@secondoftwo}%
\providecommand \href [0]{\begingroup \@sanitize@url \@href}%
\providecommand \@href[1]{\@@startlink{#1}\@@href}%
\providecommand \@@href[1]{\endgroup#1\@@endlink}%
\providecommand \@sanitize@url [0]{\catcode `\\12\catcode `\$12\catcode
  `\&12\catcode `\#12\catcode `\^12\catcode `\_12\catcode `\%12\relax}%
\providecommand \@@startlink[1]{}%
\providecommand \@@endlink[0]{}%
\providecommand \url  [0]{\begingroup\@sanitize@url \@url }%
\providecommand \@url [1]{\endgroup\@href {#1}{\urlprefix }}%
\providecommand \urlprefix  [0]{URL }%
\providecommand \Eprint [0]{\href }%
\providecommand \doibase [0]{http://dx.doi.org/}%
\providecommand \selectlanguage [0]{\@gobble}%
\providecommand \bibinfo  [0]{\@secondoftwo}%
\providecommand \bibfield  [0]{\@secondoftwo}%
\providecommand \translation [1]{[#1]}%
\providecommand \BibitemOpen [0]{}%
\providecommand \bibitemStop [0]{}%
\providecommand \bibitemNoStop [0]{.\EOS\space}%
\providecommand \EOS [0]{\spacefactor3000\relax}%
\providecommand \BibitemShut  [1]{\csname bibitem#1\endcsname}%
\let\auto@bib@innerbib\@empty
\bibitem [{\citenamefont {Feynman}\ \emph {et~al.}(1963)\citenamefont
  {Feynman}, \citenamefont {Leighton},\ and\ \citenamefont
  {Sands}}]{Feynman-lectures}%
  \BibitemOpen
  \bibfield  {author} {\bibinfo {author} {\bibfnamefont {R.~P.}\ \bibnamefont
  {Feynman}}, \bibinfo {author} {\bibfnamefont {R.~B.}\ \bibnamefont
  {Leighton}}, \ and\ \bibinfo {author} {\bibfnamefont {M.~L.}\ \bibnamefont
  {Sands}},\ }\href@noop {} {\emph {\bibinfo {title} {The {Feynman} lectures on
  physics}}}\ (\bibinfo  {publisher} {Addison-Wesley},\ \bibinfo {year}
  {1963})\BibitemShut {NoStop}%
\bibitem [{\citenamefont {Townsend}(2012)}]{Townsend-book}%
  \BibitemOpen
  \bibfield  {author} {\bibinfo {author} {\bibfnamefont {J.~S.}\ \bibnamefont
  {Townsend}},\ }\href@noop {} {\emph {\bibinfo {title} {A Modern Approach to
  Quantum Mechanics}}}\ (\bibinfo  {publisher} {University Science Books},\
  \bibinfo {year} {2012})\BibitemShut {NoStop}%
\bibitem [{\citenamefont {David~McIntyre}(2013)}]{McIntyre-book}%
  \BibitemOpen
  \bibfield  {author} {\bibinfo {author} {\bibfnamefont {J.~T.}\ \bibnamefont
  {David~McIntyre}, \bibfnamefont {Corinne A~Manogue}},\ }\href@noop {} {\emph
  {\bibinfo {title} {Quantum Mechanics}}}\ (\bibinfo  {publisher} {Pearson
  Addison-Wesley},\ \bibinfo {year} {2013})\BibitemShut {NoStop}%
\bibitem [{\citenamefont {Wilcox}(2012)}]{Wilcox-book}%
  \BibitemOpen
  \bibfield  {author} {\bibinfo {author} {\bibfnamefont {W.~M.}\ \bibnamefont
  {Wilcox}},\ }\href@noop {} {\emph {\bibinfo {title} {Quantum Principles and
  Particles}}}\ (\bibinfo  {publisher} {CRC Press},\ \bibinfo {year}
  {2012})\BibitemShut {NoStop}%
\bibitem [{\citenamefont {Mermin}(2007)}]{Mermin-book}%
  \BibitemOpen
  \bibfield  {author} {\bibinfo {author} {\bibfnamefont {N.~D.}\ \bibnamefont
  {Mermin}},\ }\href@noop {} {\emph {\bibinfo {title} {Quantum Computer
  Science}}}\ (\bibinfo  {publisher} {Cambridge University Press},\ \bibinfo
  {year} {2007})\BibitemShut {NoStop}%
\bibitem [{\citenamefont {Dur}\ and\ \citenamefont {Heusler}(2014)}]{Dur2014}%
  \BibitemOpen
  \bibfield  {author} {\bibinfo {author} {\bibfnamefont {W.}~\bibnamefont
  {D{\"u}r}}\ and\ \bibinfo {author} {\bibfnamefont {S.}~\bibnamefont {Heusler}},\
  }\href {\doibase 10.1119/1.4897588} {\bibfield  {journal} {\bibinfo
  {journal} {The Physics Teacher}\ }\textbf {\bibinfo {volume} {52}},\ \bibinfo
  {pages} {489} (\bibinfo {year} {2014})},\ \Eprint
  {http://arxiv.org/abs/https://doi.org/10.1119/1.4897588}
  {https://doi.org/10.1119/1.4897588} \BibitemShut {NoStop}%
\bibitem [{\citenamefont {D{\"u}r}\ and\ \citenamefont {Heusler}(2016)}]{Dur2016}%
  \BibitemOpen
  \bibfield  {author} {\bibinfo {author} {\bibfnamefont {W.}~\bibnamefont
  {D{\"u}r}}\ and\ \bibinfo {author} {\bibfnamefont {S.}~\bibnamefont {Heusler}},\
  }\href {\doibase 10.1119/1.4942137} {\bibfield  {journal} {\bibinfo
  {journal} {The Physics Teacher}\ }\textbf {\bibinfo {volume} {54}},\ \bibinfo
  {pages} {156} (\bibinfo {year} {2016})},\ \Eprint
  {http://arxiv.org/abs/https://doi.org/10.1119/1.4942137}
  {https://doi.org/10.1119/1.4942137} \BibitemShut {NoStop}%
\bibitem [{\citenamefont {Perry}\ \emph {et~al.}(2019)\citenamefont {Perry},
  \citenamefont {Sun}, \citenamefont {Hughes}, \citenamefont {Isaacson},\ and\
  \citenamefont {Turner}}]{Fermilab-paper}%
  \BibitemOpen
  \bibfield  {author} {\bibinfo {author} {\bibfnamefont {A.}~\bibnamefont
  {Perry}}, \bibinfo {author} {\bibfnamefont {R.}~\bibnamefont {Sun}}, \bibinfo
  {author} {\bibfnamefont {C.}~\bibnamefont {Hughes}}, \bibinfo {author}
  {\bibfnamefont {J.}~\bibnamefont {Isaacson}}, \ and\ \bibinfo {author}
  {\bibfnamefont {J.}~\bibnamefont {Turner}},\ }\href@noop {} {\bibfield
  {journal} {\bibinfo  {journal} {arXiv:1905.00282}\ } (\bibinfo {year}
  {2019})}\BibitemShut {NoStop}%
\bibitem [{\citenamefont {Shoshany}(2018)}]{Shoshany2018}%
  \BibitemOpen
  \bibfield  {author} {\bibinfo {author} {\bibfnamefont {B.}~\bibnamefont
  {Shoshany}},\ }\href@noop {} {\enquote {\bibinfo {title} {``{Thinking}
  quantum": Lectures on quantum theory},}\ } (\bibinfo {year} {2018}),\ \Eprint
  {http://arxiv.org/abs/1803.07098} {arXiv:1803.07098 [physics.pop-ph]}
  \BibitemShut {NoStop}%
\bibitem [{\citenamefont {Freericks}\ \emph {et~al.}(2019)\citenamefont
  {Freericks}, \citenamefont {Cutler}, \citenamefont {Kruse},\ and\
  \citenamefont {Vieira}}]{Freericks2019}%
  \BibitemOpen
  \bibfield  {author} {\bibinfo {author} {\bibfnamefont {J.~K.}\ \bibnamefont
  {Freericks}}, \bibinfo {author} {\bibfnamefont {D.}~\bibnamefont {Cutler}},
  \bibinfo {author} {\bibfnamefont {A.}~\bibnamefont {Kruse}}, \ and\ \bibinfo
  {author} {\bibfnamefont {L.~B.}\ \bibnamefont {Vieira}},\ }\href {\doibase
  10.1119/1.5098924} {\bibfield  {journal} {\bibinfo  {journal} {The Physics
  Teacher}\ }\textbf {\bibinfo {volume} {57}},\ \bibinfo {pages} {326}
  (\bibinfo {year} {2019})},\ \Eprint
  {http://arxiv.org/abs/https://doi.org/10.1119/1.5098924}
  {https://doi.org/10.1119/1.5098924} \BibitemShut {NoStop}%
\bibitem [{\citenamefont {Rudolph}(2017)}]{Rudolph-book}%
  \BibitemOpen
  \bibfield  {author} {\bibinfo {author} {\bibfnamefont {T.}~\bibnamefont
  {Rudolph}},\ }\href {http://www.qisforquantum.org} {\emph {\bibinfo {title}
  {Q is for Quantum}}}\ (\bibinfo  {publisher} {Terence Rudolph},\ \bibinfo
  {year} {2017})\BibitemShut {NoStop}%
\bibitem [{cee()}]{ceed}%
  \BibitemOpen
  \href {https://eng.vt.edu/ceed.html} {\emph {\bibinfo {title} {Center for the
  Enhancement of Engineering Diversity}}}\BibitemShut {NoStop}%
\bibitem [{cte()}]{ctech2}%
  \BibitemOpen
  \href {https://eng.vt.edu/ceed/ceed-pre-college-programs/c-tech2.html} {\emph
  {\bibinfo {title} {C-Tech$^2$}}}\BibitemShut {NoStop}%
\bibitem [{drq()}]{drquantumyoutube}%
  \BibitemOpen
  \href {https://www.youtube.com/watch?v=Q1YqgPAtzho} {\emph {\bibinfo {title}
  {Doctor Quantum video}}}\BibitemShut {NoStop}%
\bibitem [{\citenamefont {Kohnle}\ \emph {et~al.}(2012)\citenamefont {Kohnle},
  \citenamefont {Cassettari}, \citenamefont {Edwards}, \citenamefont
  {Ferguson}, \citenamefont {Gillies}, \citenamefont {Hooley}, \citenamefont
  {Korolkova}, \citenamefont {Llama},\ and\ \citenamefont
  {Sinclair}}]{Kohnle2012}%
  \BibitemOpen
  \bibfield  {author} {\bibinfo {author} {\bibfnamefont {A.}~\bibnamefont
  {Kohnle}}, \bibinfo {author} {\bibfnamefont {D.}~\bibnamefont {Cassettari}},
  \bibinfo {author} {\bibfnamefont {T.~J.}\ \bibnamefont {Edwards}}, \bibinfo
  {author} {\bibfnamefont {C.}~\bibnamefont {Ferguson}}, \bibinfo {author}
  {\bibfnamefont {A.~D.}\ \bibnamefont {Gillies}}, \bibinfo {author}
  {\bibfnamefont {C.~A.}\ \bibnamefont {Hooley}}, \bibinfo {author}
  {\bibfnamefont {N.}~\bibnamefont {Korolkova}}, \bibinfo {author}
  {\bibfnamefont {J.}~\bibnamefont {Llama}}, \ and\ \bibinfo {author}
  {\bibfnamefont {B.~D.}\ \bibnamefont {Sinclair}},\ }\href {\doibase
  10.1119/1.3657800} {\bibfield  {journal} {\bibinfo  {journal} {American
  Journal of Physics}\ }\textbf {\bibinfo {volume} {80}},\ \bibinfo {pages}
  {148} (\bibinfo {year} {2012})},\ \Eprint
  {http://arxiv.org/abs/https://doi.org/10.1119/1.3657800}
  {https://doi.org/10.1119/1.3657800} \BibitemShut {NoStop}%
\bibitem [{\citenamefont {Kohnle}\ \emph {et~al.}(2015)\citenamefont {Kohnle},
  \citenamefont {Baily}, \citenamefont {Campbell}, \citenamefont {Korolkova},\
  and\ \citenamefont {Paetkau}}]{Kohnle2015}%
  \BibitemOpen
  \bibfield  {author} {\bibinfo {author} {\bibfnamefont {A.}~\bibnamefont
  {Kohnle}}, \bibinfo {author} {\bibfnamefont {C.}~\bibnamefont {Baily}},
  \bibinfo {author} {\bibfnamefont {A.}~\bibnamefont {Campbell}}, \bibinfo
  {author} {\bibfnamefont {N.}~\bibnamefont {Korolkova}}, \ and\ \bibinfo
  {author} {\bibfnamefont {M.~J.}\ \bibnamefont {Paetkau}},\ }\href {\doibase
  10.1119/1.4913786} {\bibfield  {journal} {\bibinfo  {journal} {American
  Journal of Physics}\ }\textbf {\bibinfo {volume} {83}},\ \bibinfo {pages}
  {560} (\bibinfo {year} {2015})},\ \Eprint
  {http://arxiv.org/abs/https://doi.org/10.1119/1.4913786}
  {https://doi.org/10.1119/1.4913786} \BibitemShut {NoStop}%
\bibitem [{dja()}]{djalgorithm}%
  \BibitemOpen
  \href {https://en.wikipedia.org/wiki/Deutsch%E2%80%93Jozsa_algorithm} {\emph
  {\bibinfo {title} {Wikipedia page on Deutsch-Jozsa algorithm}}}\BibitemShut
  {NoStop}%
\bibitem [{\citenamefont {L{\'o}pez-Incera}\ and\ \citenamefont
  {D{\"u}r}(2019)}]{LopezIncera2019}%
  \BibitemOpen
  \bibfield  {author} {\bibinfo {author} {\bibfnamefont {A.}~\bibnamefont
  {L{\'o}pez-Incera}}\ and\ \bibinfo {author} {\bibfnamefont {W.}~\bibnamefont
  {D{\"u}r}},\ }\href {\doibase 10.1119/1.5086275} {\bibfield  {journal} {\bibinfo
   {journal} {American Journal of Physics}\ }\textbf {\bibinfo {volume} {87}},\
  \bibinfo {pages} {95} (\bibinfo {year} {2019})},\ \Eprint
  {http://arxiv.org/abs/https://doi.org/10.1119/1.5086275}
  {https://doi.org/10.1119/1.5086275} \BibitemShut {NoStop}%
\bibitem [{\citenamefont {Svozil}(2006)}]{Svozil2006}%
  \BibitemOpen
  \bibfield  {author} {\bibinfo {author} {\bibfnamefont {K.}~\bibnamefont
  {Svozil}},\ }\href {\doibase 10.1119/1.2205879} {\bibfield  {journal}
  {\bibinfo  {journal} {American Journal of Physics}\ }\textbf {\bibinfo
  {volume} {74}},\ \bibinfo {pages} {800} (\bibinfo {year} {2006})},\ \Eprint
  {http://arxiv.org/abs/https://doi.org/10.1119/1.2205879}
  {https://doi.org/10.1119/1.2205879} \BibitemShut {NoStop}%
\bibitem [{\citenamefont {Adleman}\ \emph {et~al.}(1997)\citenamefont
  {Adleman}, \citenamefont {J.DeMarrais},\ and\ \citenamefont
  {Huang}}]{adleman-qcomputability}%
  \BibitemOpen
  \bibfield  {author} {\bibinfo {author} {\bibfnamefont {L.}~\bibnamefont
  {Adleman}}, \bibinfo {author} {\bibnamefont {J.DeMarrais}}, \ and\ \bibinfo
  {author} {\bibfnamefont {M.-D.~A.}\ \bibnamefont {Huang}},\ }\href@noop {}
  {\bibfield  {journal} {\bibinfo  {journal} {SIAM J. Comput.}\ }\textbf
  {\bibinfo {volume} {26}},\ \bibinfo {pages} {1524} (\bibinfo {year}
  {1997})}\BibitemShut {NoStop}%
\bibitem [{\citenamefont {Deutsch}\ and\ \citenamefont
  {Hayden}(2000)}]{DeutschPRSLA2000}%
  \BibitemOpen
  \bibfield  {author} {\bibinfo {author} {\bibfnamefont {D.}~\bibnamefont
  {Deutsch}}\ and\ \bibinfo {author} {\bibfnamefont {P.}~\bibnamefont
  {Hayden}},\ }\href {\doibase 10.1098/rspa.2000.0585} {\bibfield  {journal}
  {\bibinfo  {journal} {Proceedings of the Royal Society of London. Series A:
  Mathematical, Physical and Engineering Sciences}\ }\textbf {\bibinfo {volume}
  {456}},\ \bibinfo {pages} {1759} (\bibinfo {year} {2000})},\ \Eprint
  {http://arxiv.org/abs/https://royalsocietypublishing.org/doi/pdf/10.1098/rspa.2000.0585}
  {https://royalsocietypublishing.org/doi/pdf/10.1098/rspa.2000.0585}
  \BibitemShut {NoStop}%
\end{thebibliography}

\end{document}